\documentclass{jfmnomark}

\usepackage{microtype}
\usepackage{graphicx}
\usepackage{natbib}
\usepackage{amsmath, mathtools}
\usepackage{tikz}
\usepackage{mathrsfs}
\DeclareMathAlphabet{\mathscrz}{OT1}{pzc}{m}{it}

\tikzset{every mark/.style={scale=0.5, color=black}}
\newcommand*\Mycirct[1]{%
  \tikz[baseline=-3pt] \draw[fill=white] (0,0) circle (4pt) node[scale=0.65] {\footnotesize \textbf{#1}};\!\!
}
\newcommand*\mycircb[1]{%
  \tikz[baseline=-1.5pt]\node[draw,circle,inner sep=0.8pt, scale=0.6] 
  {\upshape \footnotesize \textbf{#1}};\!\!
}

\newcommand*\Swtch[3]{%
  \xrightarrow[\smash{{\text{\mycircb{#2} $>$ \mycircb{#3}}}}]{\text{#1}} 
}

\def\Xint#1{\mathchoice
   {\XXint\displaystyle\textstyle{#1}}%
   {\XXint\textstyle\scriptstyle{#1}}%
   {\XXint\scriptstyle\scriptscriptstyle{#1}}%
   {\XXint\scriptscriptstyle\scriptscriptstyle{#1}}%
   \!\int}
\def\XXint#1#2#3{{\setbox0=\hbox{$#1{#2#3}{\int}$}
     \vcenter{\hbox{$#2#3$}}\kern-.5\wd0}}

\def\dashint{\Xint-}

\def\Xint#1{\mathchoice
   {\XXint\displaystyle\textstyle{#1}}%
   {\XXint\textstyle\scriptstyle{#1}}%
   {\XXint\scriptstyle\scriptscriptstyle{#1}}%
   {\XXint\scriptscriptstyle\scriptscriptstyle{#1}}%
   \!\int}
\def\XXint#1#2#3{{\setbox0=\hbox{$#1{#2#3}{\int}$}
     \vcenter{\hbox{$#2#3$}}\kern-.5\wd0}}

\def\dashint{\Xint-}
\def\ocwint{\Xint\circlearrowright}
\def\occwint{\Xint\circlearrowleft}

\newcommand{\pd}[2]{\frac{\partial #1}{\partial #2}}

\ifCUPmtlplainloaded \else
  \checkfont{eurm10}
  \iffontfound
    \IfFileExists{upmath.sty}
      {\typeout{^^JFound AMS Euler Roman fonts on the system,
                   using the 'upmath' package.^^J}%
       \usepackage{upmath}}
      {\typeout{^^JFound AMS Euler Roman fonts on the system, but you
                   dont seem to have the}%
       \typeout{'upmath' package installed. JFM.cls can take advantage
                 of these fonts,^^Jif you use 'upmath' package.^^J}%
      }
  \else
  \fi
\fi

\ifCUPmtlplainloaded \else
  \checkfont{msam10}
  \iffontfound
    \IfFileExists{amssymb.sty}
      {\typeout{^^JFound AMS Symbol fonts on the system, using the
                'amssymb' package.^^J}%
       \usepackage{amssymb}%
         
         \let\geq=\geqslant
      }{}
  \fi
\fi

\ifCUPmtlplainloaded \else
  \IfFileExists{amsbsy.sty}
    {\typeout{^^JFound the 'amsbsy' package on the system, using it.^^J}%
     \usepackage{amsbsy}}
    {\providecommand\boldsymbol[1]{\mbox{\boldmath $##1$}}}
\fi

\newcommand\hatR{\skew3\hat{R}}      % R with hat
\newsavebox{\astrutbox}
\sbox{\astrutbox}{\rule[-5pt]{0pt}{20pt}}

\newcommand\shalf{\ensuremath{{\scriptstyle\frac{1}{2}}}}

\title[New gravity-capillary waves (Part 1)]{New gravity-capillary waves at low speeds \\ Part 1: Linear geometries}

\author[P.H. Trinh and S.J. Chapman]%
{P\ls H\ls I\ls L\ls I\ls P\ls P\ls E\ns H.\ns T\ls R\ls I\ls N\ls H$^{1, 2}$\ns \and 
S.\ns J\ls O\ls N\ls A\ls T\ls H\ls A\ls N\ns C\ls H\ls A\ls P\ls M\ls A\ls
N$^2$ \break}

\affiliation{ $^1$Program in Applied and Computational Mathematics, Princeton University, \\ Washington Road,
Princeton, NJ, 08544, USA \\[\affilskip]
$^2$Oxford Centre for Industrial and Applied Mathematics, Mathematical Institute, \\ 24-29 St. Giles',
Oxford, Oxfordshire, OX1 3LB, UK}
\pubyear{1996}
\volume{538}
\pagerange{119--126}
\date{--- and in revised form ---}

\begin{document}

\maketitle

\begin{abstract}
When traditional linearised theory is used to study gravity-capillary waves produced by flow past an obstruction, the geometry of the object is assumed to be small in one or several of its dimensions. In order to preserve the nonlinear nature of the obstruction, asymptotic expansions in the low-Froude or low-Bond number limits can be derived, but here, the solutions invariably predict a waveless surface at every order. This is because the waves are in fact, exponentially small, and thus \emph{beyond-all-orders} of regular asymptotics; their formation is a consequence of the divergence of the asymptotic series and the associated Stokes Phenomenon.  

By applying techniques in exponential asymptotics to this problem, we have discovered the existence of new classes of gravity-capillary waves, from which the usual linear solutions form but a special case. In this paper, we present the initial theory for deriving these waves through a study of gravity-capillary flow over a linearised step; this will be done using two approaches: in the first, we derive the surface waves using the the standard method of Fourier transforms; in the second, we derive the same result using exponential asymptotics. Ultimately, these two methods give the same result, but conceptually, they offer different insights into the study of the low-Froude, low-Bond number problem. 
\end{abstract}


\section{Introduction}

\noindent Our interest in the gravity-capillary problem lies with
the complicated scenario of waves induced by interactions with objects in a stream. Here, the first significant theory was
proposed by Lord Rayleigh \citeyearpar{rayleigh_1883}, whose chief source of
inspiration had been the earlier experiments of Scott Russell
\citeyearpar{russell_1844} and Thomson \citeyearpar{thomson_1871}. As \cite{rayleigh_1883} himself explains:
\begin{quotation} 
\noindent  \emph{When a small obstacle, such as a fishing line, is [...] held stationary in moving water, the surface is covered with a beautiful wave-pattern, fixed relatively to the obstacle. On the up-stream side the wavelength is short, and,
as Thomson has shown, the force governing the vibrations is principally
cohesion. On the down-stream side the waves are longer, and are governed
principally by gravity.} 
\end{quotation}

\noindent Rayleigh supposed that the fluid could be assumed to be
two-dimensional, inviscid, incompressible, and irrotational, and that the
effects of the fishing line could be approximated by the application of a
small pressure distribution to a single point on the free-surface (as might be
produced by a small jet of air). Then, by linearising for small-amplitude waves
and for a weakly-applied pressure distribution, the method of Fourier
transforms produces an approximation of the upstream capillary waves and
downstream gravity waves. As Rayleigh remarked, the theory is particularly
successful in predicting that those particular wave patterns can only exist when
the speed of the stream is somewhat faster than $23$ centimetres per second.

\begin{figure}
\includegraphics[width=1.0\linewidth]{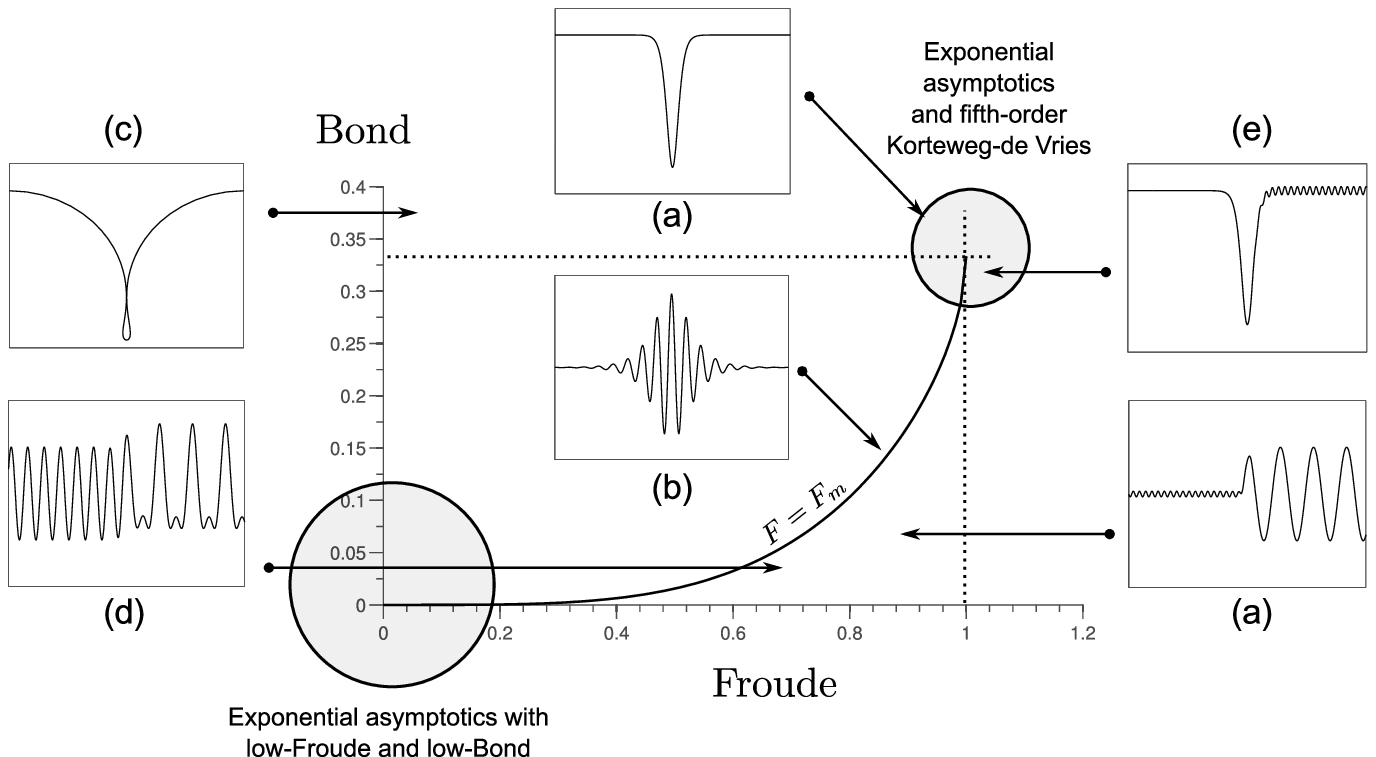}
\begin{caption}{%
Many types of gravity-capillary waves can be found when studying
steady finite-depth flows subject to a disturbance. Possible solutions include profiles with
(a) trapped bubbles, (b) solitary waves, (c) generalised solitary waves, (d)
capillary waves upstream and gravity waves downstream, (e) localised
wavepackets, and (f) Wilton ripples. The circled region in the lower-left,
corresponding to low-Froude and low-Bond flows has been largely ignored and we
aim to study this asymptotic limit. \label{fig:gclin_gctotal}}
\end{caption}
\end{figure}

Since Lord Rayleigh's seminal work, however, mathematical analyses of the
general gravity-capillary problem have led to the realisation that, while
the study of gravity-only or capillary-only flows are themselves rich in
difficulties, the combination of both effects presents a much more formidable
challenge. For the theory of finite-depth gravity-capillary waves, the Froude number $F$, 
representing the ratio between inertial and gravitational forces, and the Bond
number $B$, representing the ratio between gravitational and surface tension
forces, are the two crucial parameters in the problem. In Figure
\ref{fig:gclin_gctotal}, we have illustrated a few of the different kinds of solutions that might be expected when studying even the simplest problems incorporating both effects. Generally, we can expect qualitatively similar types of behaviours even for different types of geometries (finite or infinite depth, perturbed by a submerged object or a pressure distribution, \emph{etc.}). Thus, we will continue to refer to Figure \ref{fig:gclin_gctotal}, even for the case of Rayleigh's solution (for example), which is only technically applicable for infinite-depth flows.

The figure contains several key regions of interest that are primarily separated by the curve $F = F_m$, the line $B = 1/3$, and the line $F = 1$; we will address these bifurcations later in the text, and they are also described in Appendix \ref{sec:gclin_class}. Notice that if we begin with Rayleigh's fishing line problem, then solutions like the one in (d) may appear, but only in the region below the critical \emph{dispersion curve} of $F=F_m$. In addition to Rayleigh's work, \cite{forbes_1983} and \cite{grandison_2006} have also studied this problem of capillary-upstream, gravity-downstream waves, but for flows over a semi-circular obstruction. Within the region below $F = F_m$, it is also possible to find solutions (f) exhibiting the so-called \emph{Wilton Ripples} phenomenon \citep{wilton_1915}; solutions with these secondary ripples have been computed numerically for the waves generated by a moving pressure distribution \citep{vb_2002}, and have also been seen in nature \citep{schooley_1960}. 

If the Froude number is now held fixed, and the Bond number is increased, then as we approach the critical dispersion curve, the solution set bifurcates. In particular, Rayleigh's result is singular as this curve is approached, and the system may bifurcate into one of many possible solutions in the form of envelope solitary waves (e), where an appropriate model is the forced Nonlinear Schr\"{o}dinger Equation. If the Bond number is increased further and past $B = 1/3$, then the solutions take the form of classical solitary waves (b) and can then be approximated by a forced Korteweg-de Vries Equation. Within all these regions, there is occasionally the possibility of non-uniqueness, and a single point in $(F,B)$ space often does corresponds to multiple solutions, for which only one represents a perturbation from the uniform flow; for example, solutions can exhibit trapped bubbles (a), representing a perturbation from Crapper's \citeyearpar{crapper_1957} exact solution of the unforced pure-capillary problem. These issues have been investigated for the case of flow induced by an applied pressure distribution by many authors, including \cite{vb_1992}, \cite{dias_1993}, \cite{dias_1996}, and \cite{maleewong_2005, maleewong_2005b}.

Finally, the historically controversial region just to the right of the line $F = 1$ separates subcritical and supercritical flows, and exists as a warning sign to those who remain unaware of the importance of exponentially small terms in singular problems. The earliest numerical
simulations of this region, computed from the full nonlinear equations, were performed by \cite{hunter_1983}. What they discovered was that when the Bond number is greater than $1/3$, the solution ressembles the classical depression solitary wave of (b); but if the Bond number is instead less than $1/3$, the numerical solutions are bedeviled by small-scale oscillations near the tails (c). These dimpled solutions could not be explained at the time, and their significance was quickly glossed over. 

However, in the years that followed, a series of papers by
\cite{hunter_1988}, \cite{pomeau_1988}, \cite{beale_1991}, and \cite{sun_1991}
established the notion of nonlocal (or generalised) solitary
waves---essentially, a solitary wave coupled with exponentially small
oscillations near the tails. Near the point $F = 1$, $B = 1/3$, the water waves
are governed by the fifth-order Korteweg-de Vries Equation, for which recent
standardised techniques in exponential asymptotics can be applied
\citep{grimshaw_2010, trinh_2010}. Indeed such ripples have been observed for the related case of internal gravity waves \cite{akylas_1992}. The unique history and ultimately,
the resolution of the fifth-order Korteweg-de Vries Equation is well documented
and described in \cite{boyd_wnlsw}.

\subsection*{What do we plan to accomplish?}

There are two areas that still require much work: what is the nature of gravity-capillary waves over more general topographies that \emph{cannot} be considered small (such as over a large step in a channel or past any full-bodied obstruction) and in particular, what analytical and numerical work can be done in the regime where the Froude and Bond numbers tend to zero.

The crucial idea is that in the zero-Froude and zero-Bond solution, the
free-surface is replaced by a rigid wall, with the full geometry of the
obstruction preserved in the solution. Thus, asymptotic approximations for
low-Froude and low-Bond numbers allow us to directly relate the
generation of waves to the shape of the obstruction. This is in marked contrast
to more traditional linear approximations which depend on an asymptotic limit of
small obstructions.  The difficulty in the low-Froude, low-Bond limits, however, is that it represents a singular limit. At ever-decreasing Froude and Bond numbers, we
would still expect waves to appear on the free-surface, but it is easy to check
that the approximation at every single order in the asymptotic hierarchy yields
a waveless solution. The waves are, in fact, exponentially small and thus
\emph{beyond-all-orders} of traditional asymptotics; in the context
of gravity-only flows, this has been termed the \emph{Low-Speed Paradox}, first
mentioned by \cite{ogilvie_1968}.

In this paper, we will study the problem of flow over a rectangular step, but
rather than performing the full low-Froude, low-Bond analysis for
$\mathcal{O}(1)$ geometries, we first linearise for asymptotically small steps.
The mathematical formulation is presented in \S\ref{sec:gclin_form}, and from
here, the analysis can proceed in two equivalent ways. In
\S\ref{sec:gclin_fourier}, we will regard the Froude and Bond numbers as fixed
and $\mathcal{O}(1)$, and solve using Fourier transforms, after which we take
$F, B \to 0$. This is to be contrasted to \S\ref{sec:gclin_expasym}, where we
will take $F, B \to 0$ immediately after the initial linearisation, and then
make use of exponential asymptotics. 

Our goals are twofold. First, we wish to compare and contrast the
method of Fourier transforms with the method of exponential asymptotics. Do
concepts like the Stokes Phenomenon and optimal truncation become
more transparent using one approach over the other? Second, we wish to show how exponential asymptotics can be applied to the simpler linearised equations of the small-step problem. In the continuation of this paper (Part 2), we will work with the full nonlinear equations; this first excursion, then, provides a smooth transition to the more advanced techniques that underly the subsequent work. Similar ideas of comparing and contrasting the Fourier and exponential-asymptotics approach were used by \cite{mortimer_thesis} and \cite{chapman_2005} in the study of partial differential equations. 

\section{Illustration of the general methodology} \label{sec:gclin_method}

\noindent Before proceeding to the mathematical analysis, however, we shall describe the key ideas which underly the study to come.

Consider flow over a rectangular step in a channel. From the
introduction, we know that for subcritical ($F < 1$) flows linearised theory
predicts two regions of interest. In Region I below the critical dispersion
curve of Figure \ref{fig:gclin_disproots}, we expect flows with capillary waves
upstream and gravity waves downstream, while in Region II above the
dispersion curve, we expect localised solitary waves. As we will show in
\S\ref{sec:gclin_fourier}, the key difference between Type I and II solutions is
the kind of residue contributions that are collected in the Fourier inversion
process: Region I solutions have four real residues (representing two real
wavenumbers) while Region II solutions have four complex residues. However, this
phenomenon can also be interpreted using exponential asymptotics, and we now
give a taste of the underlying ideas of \S\ref{sec:gclin_expasym}. 

When the free surface, described by the fluid speed $q(\phi)$, and streamline
angle $\theta(\phi)$, as functions of the potential $\phi$, is expanded into an
asymptotic series in powers of the Froude and Bond numbers, then we find that
the expansions of $q$ and $\theta$ are waveless to every order. The
exponentially small waves on the free-surface must instead be interpreted as
arising from the \emph{Stokes Phenomenon}, by which the coefficient of a subdominant exponential in an asymptotic expansion appears to change discontinuously as a so-called \emph{Stokes line} is crossed. These concepts are connected to the divergence of the asymptotic
expansions, and are reviewed in, for example, the works of \cite{dingle_book}
and \cite{boyd_wnlsw, boyd_1999}. 

\begin{figure}
\includegraphics{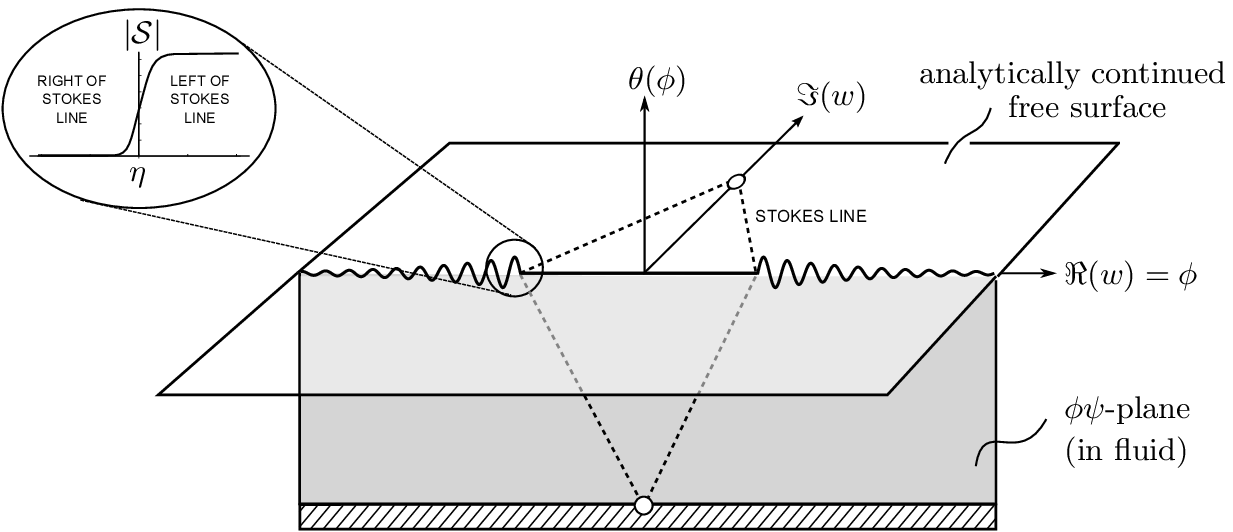}
% \begin{preview}
%  %\begin{overpic}[scale=0.50, grid, tics=10]{figpdf/switchplot.pdf}
%  \begin{minipage}{1.0\textwidth}
%  \vspace*{0.2cm}
%  \begin{overpic}[scale=0.44]{figpdf/switchplot2.pdf}
%  \put(95,22) {$\Re(w) = \phi$}
%  \put(56,39.5) {$\theta(\phi)$}
%  \put(11.4,31.2) {\footnotesize $\eta$}
%  \put(10.8,43.6) {\footnotesize $\lvert \mathcal{S} \rvert$}
%  \put(71,38) {$\Im(w)$}
%  \put(95,10) {\footnotesize $\phi\psi$-plane}
%  \put(95, 6.3) {\footnotesize (in fluid)}
%  \put(82, 40) {\footnotesize analytically continued}
%  \put(90, 37) {\footnotesize free surface}
%  \end{overpic}
%  \end{minipage}
%  \end{preview}
\caption{%
The \emph{physical} solutions of interest is $\theta$, the angle of the free-surface as a function of the potential $\phi\in\mathbb{R}$. The free-surface can be analytically continued, producing the perpendicular plane, $\phi = w \in \mathbb{C}$. This complex plane shares a correspondance with the physical $\phi\psi$-plane (darkly shaded). Singularties are shown as circles and Stokes lines are dashed. Singularites and Stokes lines in the lower complex $w$-plane are not illustrated. \label{fig:gclin_method}} 
\end{figure}

Of course, the velocity $qe^{-i\theta}$ is entirely well-behaved on the
free surface, but its analytic continuation contains a singularity
corresponding to the sharp rise in the step. It is this singularity that causes the asymptotic series to diverge, and is the origin of the Stokes lines. To study this issue, we complexify the free-surface, sending $\phi \mapsto \overline{\phi} + i\overline{\psi} = \overline{w} \in \mathbb{C}$ and $qe^{-i\theta} \mapsto \overline{q}e^{-i\overline{\theta}} = d\overline{w}/d\overline{z} \in \mathbb{C}$. This is represented by the perpendicular plane (out of the page) in Figure \ref{fig:gclin_method}. However, because of the nature of complex variables, we may continue to identify these new complexified variables, $\overline{w}$ and $d\overline{w}/d\overline{z}$, with the usual quantities of $w$ (the complex potential) and $dw/dz$ (the complex velocity). 

In Figure \ref{fig:gclin_method}, the darkly shaded plane depicts the \emph{physical} $\phi\psi$-plane---that is, the plane where fluid lies above the solid boundary, and where there is a single singularity due to the small step. In this work, we may speak of the \emph{corner of the step generating a wave}, but we are actually referring to the singularity in the analytic continuation, identifiable with the step, rather than the step itself.

Once we have located these singularities, then we can expect Stokes lines (dashed in Figure \ref{fig:gclin_method}) to emerge from each one, across which the Stokes Phenomenon necessitates the switching-on of exponentially small waves (whose amplitudes scales like $|\mathcal{S}|$ in Figure \ref{fig:gclin_method}). Afterwards, by the process of optimal truncation and Stokes smoothing, we can explicitly derive the form of the waves that switch on. Our work in this paper is an extension of the methods developed by \cite{chapman_2002} for capillary waves, and \cite{chapman_2006} for gravity waves. These studies are themselves based on the procedure of optimal truncation and Stokes smoothing, as described in \cite{olde_1995} and \cite{chapman_1998}. In fact, the rectangular step in this paper is chosen for its simplicity; the exponential asymptotics theory will be later applied (in Part 2) to flows over more general nonlinear geometries (see also related work by \cite{trinh_1hull} and \cite{lustri_2012}]. 

It is important to note that this technique of optimal truncation and Stokes smoothing is presented here in preparation for the nonlinear analysis of Part 2. Indeed, for the simpler problem of Part 1, it is an equally valid approach to apply a WKB analysis to the linearised equation and search for solutions of the form $Qe^{-\chi/\epsilon}$. In the nonlinear variant, the same idea allows us to determine the argument of the exponentials, $\chi$, but not the constant (Stokes) multipliers, $Q$. 
Finding these Stokes multipliers is the key difficulty of nonlinear problem, and one method is to use the exponential asymptotics presented here. 

However, the methodology we use is not the only such approach for studying exponentially phenomena. Indeed, there exists a substantial community that uses, for example, Borel summation and for which the relevant waves originate from singularities in the Borel plane. We refer readers to the tutorials and reviews by Boyd (\citeyear[Chap. 4]{boyd_wnlsw}), \cite{olde_1999}, \cite{costin_book}, \cite{grimshaw_2010} and the references therein.

\section{Mathematical formulation} \label{sec:gclin_form}

\begin{figure}
\includegraphics{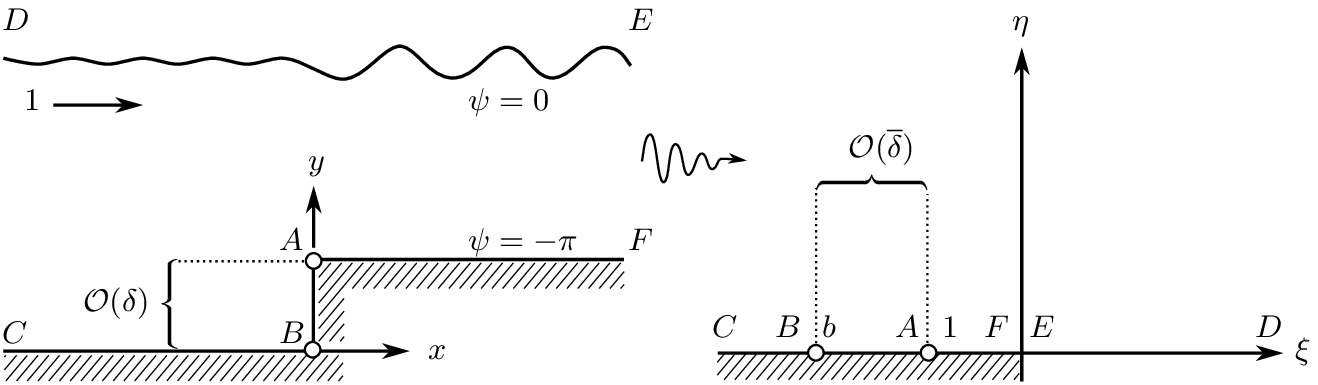}
% \begin{preview}
% \begin{minipage}{1.0\textwidth}
%  \vspace*{0.1cm}
%  %\begin{overpic}[scal1e=0.38, grid, tics=10]{figpdf/form.pdf}
%  \begin{overpic}[scale=0.37]{figpdf/form.pdf}
%  \put (35,10){\small $\psi = -\pi$}
%  \put (35,20.5){\small $\psi = 0$}

%  \put (97,1.8){\small $\xi$}
%  \put (75.8,26.5){\small $\eta$}
 
%  \put (23,16){\small $y$}
%  \put (32,1.8){\small $x$}
%  \put (6.1,5.4){\small $\mathcal{O}(\delta)$}
%  \put (1.7,20.5){\small $1$}
 
%  \put (20.8,10){\small $A$}
%  \put (20.8,3){\small $B$}
%  \put (0,3){\small $C$}
%  \put (0,26.5){\small $D$}
%  \put (47,26.5){\small $E$}
%  \put (47,10){\small $F$}
 
%  \put (63.5,17){\small $\mathcal{O}(\overline{\delta})$}
 
%  \put (67,3.5){\small $A$}
%  \put (70.5,3.5){\small $1$}
%  \put (58,3.5){\small $B$}
%  \put (61.5,3.5){\small $b$}
 
%  \put (53.3,3.5){\small $C$}
%  \put (94,3.5){\small $D$}
%  \put (77.1,3.5){\small $E$}
%  \put (73.7,3.5){\small $F$}
%  \end{overpic}
%  \end{minipage}
% \end{preview}
\caption{%
The flow in the physical $xy$-plane (left) is first mapped to the
potential $w = \phi + i\psi$ plane, then again mapped to the upper half $\zeta =
\xi + i\eta$ plane (right) using $\zeta = e^{-w}$. Later, the analysis will proceed by assuming that the height of the step is small. \label{fig:gclin_form}}
\end{figure}

\noindent Consider steady two-dimensional potential flow of an incompressible
fluid over a rectangular step in a channel (Figure \ref{fig:gclin_form}). Far upstream, the flow is uniform with constant velocity $U$ and the height of the channel is $L$. We first non-dimensionalise the velocity with $U$ and the length with $L/\pi$. Then, if the fluid velocity is $\textbf{u} = (u, v) = \nabla \phi$, the potential satisfies Laplace's equation,
\begin{equation}
 \nabla^2 \phi = 0.
\end{equation}

\noindent On all boundaries, we have the kinematic condition,
\begin{equation}
 \pd{\phi}{n} = 0,
\end{equation}

\noindent and on the free surface, Bernoulli's equation gives
\begin{equation} \label{eq:gclin_dyn_a}
 \frac{F^2}{2}\left( \left| \nabla\phi\right|^2 - 1\right) + y =
-B\kappa,
\end{equation}

\noindent where $F = U/\sqrt{gL}$ is the Froude number defined for gravitational
constant $g$; $B = \sigma/(\rho g L^2)$ is the Bond number defined according to
the surface tension parameter $\sigma$ and density $\rho$; and $\kappa$ is the
curvature, defined to be positive if the center of curvature lies in the fluid
region. 

We then define the complex potential by $w = \phi + i\psi$, where $\psi$ is the streamfunction, and conformally map the flow from the physical plane to the strip
between $\psi = -\pi$ and $\psi = 0$. A second transformation with $w \mapsto \zeta = \xi + i\eta = e^{-w}$ is then used to map the potential plane to the
upper-half $\zeta-$plane. The stagnation and corner points of the step
are mapped to $\zeta = -1$ and $\zeta = -b$, respectively, and we can prescribe
the step by $y = \delta f(x) = \delta H(x)$ where $H(x)$ is the Heaviside
function and $\delta \ll 1$. These physical and $\zeta$-planes are shown in
Figure \ref{fig:gclin_form}.

The flow can now be handled more easily by formulating
the problem in terms of the hodograph variables, $\log\left(dw/dz\right) =
\log{q} - i\theta$, where $q$ is the speed of the flow and $\theta$ is the angle
the streamines make with the $x$-axis. Then by Cauchy's Theorem, it can be shown that
\begin{equation} \label{eq:gclin_bdint_re}
 \log{q} = -\frac{1}{\pi}\dashint_{-\infty}^\infty \frac{\theta(\xi')}{\xi'-\xi}
\ d\xi'.
\end{equation}

\noindent Reposing the potential problem in terms of this integral formulation is a standard technique---see for example, \cite{king_1987, king_1990} and \cite{tuck_1991} for further descriptions. Bernoulli's equation (\ref{eq:gclin_dyn_a}) can also be transformed into a more convenient form by differentiating with respect to the arclength parameter $s$ (increasing in the flow direction), and using
\[
 \frac{d}{ds} = \frac{d\phi}{ds}\frac{d}{d\phi} = q \frac{d}{d\phi}.
\]

\noindent Then after rearranging, Bernoulli's equation becomes
\begin{equation} \label{eq:gclin_dyn_re}
 F^2 \left[ q^2 \frac{dq}{d\phi} \right]
- B \left[ q^2 \frac{d^2\theta}{d\phi^2} +
 q\frac{dq}{d\phi}\frac{d\theta}{d\phi} \right] =
- \sin\theta.
\end{equation}

\noindent Note that in the usual treatment of linearised flow in a channel
(\emph{cf}. \citealt{lamb_book} and \citealt{king_1987,king_1990}) the problem is
typically nondimensionalised so that the channel depth is $1$. Our decision to 
nondimensionalise with a depth of $\pi$ is instead consistent with the
choice of \cite{chapman_2006}.

\section{Classical linearised theory for small steps} \label{sec:gclin_fourier}

\noindent Linearised solutions for flow over a slight bottom topography have
been treated for the case of gravity waves in the classic text of
\cite{lamb_book} as well in \cite{king_1987, king_1990}. For the case of
gravity-capillary waves, \cite{forbes_1983} performed the analysis for flow
over a semi-circular obstruction (using the Joukowski map), but surprisingly,
a similar analysis does not seem to have been done for the case of flow over a
step.

First, so as to distinguish the free-surface and boundary, we use the
following notation: if $\xi > 0$ and hence we are positioned on the
free-surface, then we let $\xi = e^{-\phi}$ and $\theta(\xi)= \theta_f (\phi)$,
where $\phi$ varies between $\infty$ and $-\infty$; otherwise, if $\xi < 0$ and
we are positioned on the channel bottom, then we let $\xi = -e^{-\phi}$ and
$\theta(\xi) = \theta_b(\phi)$, where now $\phi$ varies between
$-\infty$ and $\infty$. The boundary integral (\ref{eq:gclin_bdint_re}) can first be written as 
\begin{align*}
 -\frac{1}{\pi} \dashint_{-\infty}^\infty \frac{\theta(\xi')}{\xi' - \xi} d\xi'
 &= - \frac{1}{\pi} \Biggl[ \dashint_{-\infty}^\infty
\frac{\theta_{f}(s)}{1 - e^{-(\phi-s)}} ds - \int_{-\infty}^\infty
\frac{\theta_{b}(t)}{1 + e^{-(\phi-t)}} dt \Biggr] \\
& \equiv P_f(\phi)
\end{align*}

\noindent with $\phi \in (-\infty, \infty)$. $P_f(\phi)$ can then be written
as a sum of convolutions
\begin{equation}
 P_f(\phi) = - \frac{1}{\pi}\biggl[ (\theta_f \ast k_+)(\phi)
- (\theta_b \ast k_-)(\phi) \biggr],
\end{equation}

\noindent where
\begin{equation}
 k_\pm(\sigma) = \frac{1}{1 \mp e^{-\sigma}},
\end{equation}

\noindent and the convolution product is defined by
\begin{equation}
 (g \ast h)(\phi) \equiv \int_{-\infty}^\infty g(s) \cdot h(\phi - s) \ ds.
\end{equation}

\noindent To linearise $\theta$ for small steps, we set
\begin{equation}
 \theta = \delta \theta^{(1)} + \delta^2
\theta^{(2)} + \ldots
\end{equation}

\noindent so that for $\xi > 0$ and along the free-surface, $q(\xi) = \exp
\left\{P_f(\phi) \right\} = 1 + \mathcal{O}(\delta)$ from the expression in 
(\ref{eq:gclin_bdint_re}). Then at $\mathcal{O}(\delta)$ we get:
\begin{gather}
\overline{\theta}_b^{(1)} = f'(\phi), \label{eq:gclin_bdintfirst} \\
F^2 \frac{dP_f^{(1)}}{d\phi} - B \frac{d^2
\theta_f^{(1)}}{d\phi^2} = -\theta_f^{(1)}. \label{eq:gclin_dynfirst}
\end{gather}

\noindent Here and henceforth, primes denote differentiation in $\phi$ (or
later, $w$). Note that (\ref{eq:gclin_bdintfirst}) can be obtained using the
relations $\tan \theta_b = f'(x(\phi))$ and $dx/d\phi = \cos\theta/q$, applied
along the channel bottom. Finally, (\ref{eq:gclin_dynfirst}) can be written as 
\begin{equation} \label{eq:gclin_dynconv}
- \frac{F^2}{\pi} \frac{d}{d\phi}\biggl[ (\theta \ast k_+)(\phi) -
(f' \ast k_-)(\phi) \biggr] - B \frac{d^2 \theta}{d\phi^2} =
-\theta.
\end{equation}

\noindent In this last equation, we have replaced the notation of
$\theta_f^{(1)}$ with $\theta$, so long as it remains clear that we are only
interested in the first approximation, evaluated along the free-surface. 

\subsection{Fourier integrals}

\noindent We will define the Fourier transform of a function, $f(x)$ as
\[
 \mathscr{F}[f] \equiv \int_{-\infty}^\infty f(x)e^{-ikx} \ dx.
\]
 
\noindent Taking the transform of eqn (\ref{eq:gclin_dynconv}) gives
\begin{align*}
-\frac{F^2}{\pi} (-ik) \biggl[ \mathscr{F}[\theta]
\mathscr{F}[k_+] - \mathscr{F}[f'] \mathscr{F}[k_-] \biggr] - B(ik)^2
\mathscr{F} [\theta] &= -\mathscr{F}[\theta],
\end{align*}

\noindent and we can use the fact that
\begin{equation}
 \mathscr{F}[k_+] = -\frac{\pi i}{\tanh \pi k} \text{\ \quad and \ \quad} \mathscr{F}[k_-] = -\frac{\pi i}{\sinh \pi k},
\end{equation}

\noindent and rearrange to get
\begin{equation}
\mathscr{F}[\theta] = \frac{k F^2}{kF^2 \cosh \pi k -
\sinh \pi k (Bk^2 + 1)} \int_{-\infty}^\infty f'(t) e^{-ikt} \ dt.
\end{equation}

\noindent Inversion then gives the form of the free-surface:
\begin{equation} \label{eq:gclin_theta_compact}
\theta = \frac{F^2}{2\pi}\int_{-\infty}^\infty \frac{k}{g(k)\cosh
\pi k } \left[ \int_{-\infty}^\infty f'(t) e^{ik(t-\phi)} \ dt \right] dk,
\end{equation}

\noindent where
\begin{equation} \label{eq:gclin_gk}
g(k) = k F^2 - \tanh (\pi k) (Bk^2 + 1)
\end{equation}

\noindent is the dispersion relation and we will go on to define the
inversion contour more specifically in the next section. The different
root arrangements of the dispersion relation and their
relationships with the various solutions mentioned in the
introduction is expounded in Appendix \ref{sec:gclin_class}, but
briefly, there are two types of poles: the first type lies entirely on the
imaginary axis and affects the free surface primarily near the origin $\phi = 0$.
Setting $k = \pm \beta_0, \pm \beta_1, \ldots$, where $\beta_i \in \mathbb{R}$,
we see that these poles are given by solving the equation:
\begin{equation} \label{eq:gclin_Bn_eqn}
 \frac{\tan (\pi \beta_n)}{\beta_n} = \frac{F^2}{1 - B \beta_n^2}, \qquad \text{for $n = 1, 2, 3, \ldots$}
\end{equation}

\noindent and where $\beta_0 = 0$. The second type of pole corresponds to the
wavenumbers of the gravity-capillary waves; if we let $k = \pm k_0, \pm
k_1$ be these wavenumbers and order them according to their behaviours as
determined in Region I (of Figure \ref{fig:gclin_disproots}), then $\pm
k_0$ corresponds to gravity waves and $\pm k_1$ to capillary waves, with $0< k_0
< k_1$. However, if the Froude and Bond numbers are chosen to lie in Region II (of Figure \ref{fig:gclin_disproots}), then the four residues will be entirely
complex, and we identify the poles in the lower-half plane with $k_0$ and
$-\overline{k_0}$, and the poles in the upper-half plane with $k_1$ and
$-\overline{k_1}$. The reasons for this will become clear in the next section.

%Lastly, note the following properties of $g(k)$ and $g'(k)$, which will be
%useful in the sections to follow:
%\begin{align}
%g(-k) &= -g(k) & g'(-k) &= g'(k) \label{eq:gclin_gprop1} \\
%g(\overline{k}) &= \overline{g(k)} & g'(\overline{k}) &= \overline{g'(k)} 
%\label{eq:gclin_gprop2}
%\end{align}

\subsection{Fourier inversion} \label{sec:gclin_fourinv}

\noindent At this point, we will now substitute the expression $f(t) = H(t)$
for the geometry of the step into (\ref{eq:gclin_theta_compact}), giving
\begin{equation} \label{eq:gclin_theta_compact2}
\theta = \frac{F^2}{2\pi}\int_{-\infty}^\infty \frac{k e^{-ik\phi}}
{g(k)\cosh (\pi k) } \ dk.
\end{equation}

\begin{figure}\centering
% \beginpgfgraphicnamed{gclinear_intcontour}
% \begin{tikzpicture}
% \node at (0,0){\includegraphics[width=0.93\linewidth]{figpdf/intcontour.pdf}};
% \node at (-2,-0.45){$-k_0$};
% \node at (1.4,-0.45){$k_0$};
% \node at (3.25,0.3){$k_1$};
% \node at (-4,0.3){$-k_1$};
% 
% \node at (6.5, -0.1){$\Re(k)$};
% \node at (-0.2,3.7){$\Im(k)$};
% \node at (3.6,3.50){$\phi < 0$};
% \node at (3.6,-3.70){$\phi > 0$};
% 
% \node at (-0.8,0.3){$i\beta_1$};
% \node at (-0.8,0.8){$i\beta_2$};
% \node at (-0.8,1.3){$i\beta_3$};
% \node at (-0.8,1.8){$i\beta_4$};
% 
% \node at (-0.9,-0.6){$-i\beta_1$};
% \node at (-0.9,-1.1){$-i\beta_2$};
% \node at (-0.9,-1.6){$-i\beta_3$};
% \node at (-0.9,-2.1){$-i\beta_4$};
% 
% \node at (2.9,-1.8){$k_0$};
% \node at (2.9, 1.6){$k_1$};
% \node at (-3.6,-1.8){$-\overline{k_0}$};
% \node at (-3.6,1.6){$-\overline{k_1}$};
% \end{tikzpicture}
% \endpgfgraphicnamed
\includegraphics[width=1.0\textwidth]{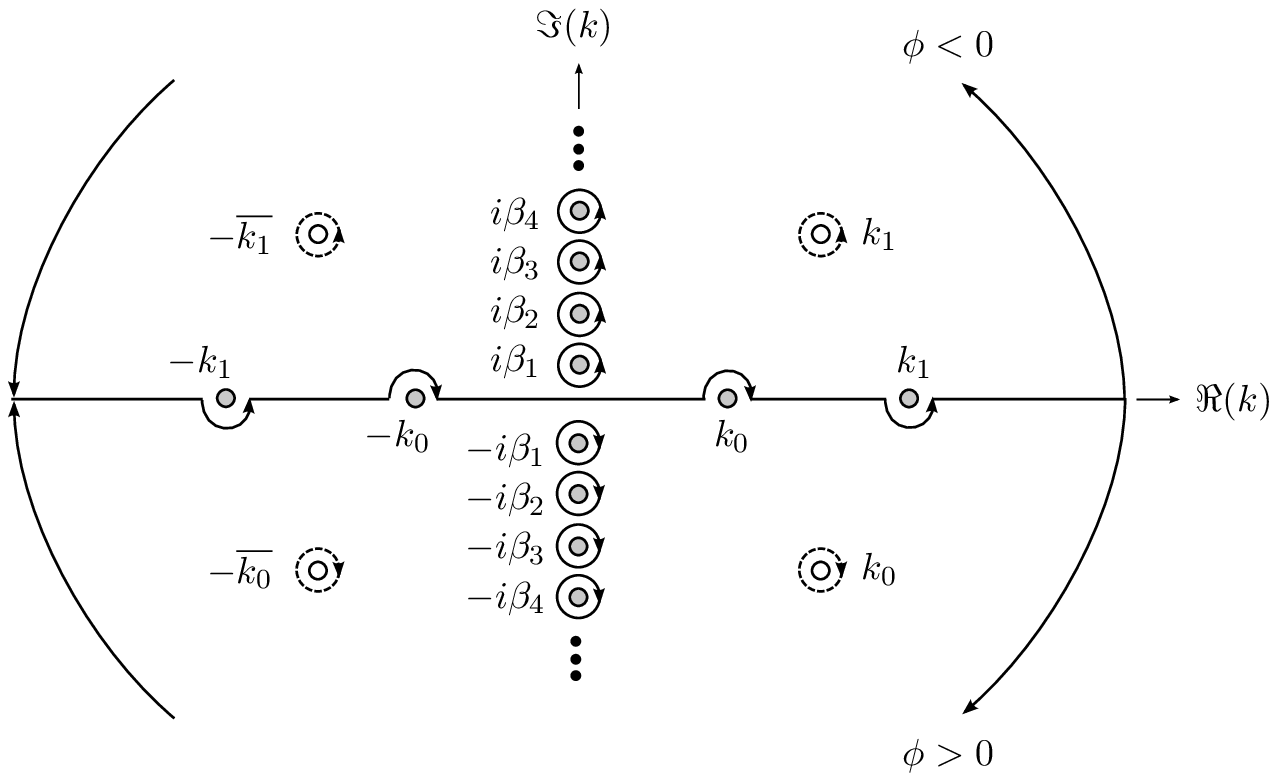}
\caption{%
When there are four real wavenumbers, the Fourier inversion of eqn
(\ref{eq:gclin_theta_compact2}) is done so that for $\phi > 0$, the gravity
residues ($k = \pm k_0$) are collected, whereas for $\phi < 0$, the
capillary residues ($k = \pm k_1$) are collected. If there are
four complex wavenumbers, then the residues in the upper and lower-half plane
are similarly collected, in the orientation prescribed by the dotted lines.\label{fig:gclin_fourinv}}
\end{figure}

\noindent First, let us perform the Fourier inversion for solutions in Region I, so with four real wavenumbers; the contour is illustrated in Figure \ref{fig:gclin_fourinv}. 
 
The definition of the integral in (\ref{eq:gclin_theta_compact2}) suffers from
the presence of the four poles along the real axis, and this is a symptom of the
unnatural and idealised steady-state problem [\emph{c.f.} Stoker (\citeyear{stoker_book}, p.174) and Debnath (\citeyear{debnath_book}, p.44)]. We must re-define the integration contour along the real $k$-axis, taking in account the
correct radiation conditions far upstream and far downstream from the
disturbance. Suppose that $\phi > 0$: in this case, we deform the contour
into the lower-half $k$-plane, where the integrand tends to zero as $k \to
-i\infty$. In the limit that $\phi \to \infty$, we want to preserve only gravity
waves, and so we only collect the residues from $k = \pm k_0$. Similarly, if
$\phi < 0$, then we still deform into the upper-half $k$-plane, but the path is
chosen so that the residues $k = \pm k_1$ are collected as $\phi \to -\infty$. 

With this in mind, we may write the Fourier integral (\ref{eq:gclin_theta_compact2}) as
\begin{equation} \label{eq:gclin_fourinv}
\theta =
\frac{F^2}{2\pi} \int_{-\infty}^\infty 
\frac{G(k)}{g(k)} \ dk  = \frac{F^2}{2\pi}
\begin{dcases}
\left[\ocwint_{+k_0} + \ocwint_{-\overline{k_0}} + \sum_{n=1}^\infty \ocwint_{-i \beta_n} 
\right] \frac{G(k)}{g(k)} \ dk & \quad
\phi > 0 \\
\left[ \occwint_{+k_1} + \occwint_{-\overline{k_1}} + \sum_{n=1}^\infty \occwint_{+i
\beta_n} \right] \frac{G(k)}{g(k)} \ dk &
\quad \phi < 0,
\end{dcases}
\end{equation}

\noindent where $G(k) = k e^{-ik \phi}/\cosh (\pi k)$ and the integral-shorthand represents contour integrals around the indicated poles. Note that we have also included the residue contribution from $k = i\beta_0 = 0$ since this
is simply zero. Rather than performing the separate cases for the upstream and downstream solutions, let us write $k = k^*$ for either $k_0$ or $k_1$, and where we thus require $\Re(k^*) > 0$ and $\Im(k^*) \gtrless 0$ for $\phi \lessgtr 0$. 

Away from the critical curves and between regions in Figure \ref{fig:gclin_disproots}, the poles are simple roots of $g(k)$, so the upstream and downstream solutions can be written as
\begin{align}
\theta_\pm &= \frac{F^2}{2\pi} \left[ \pm 2 \pi i
\left\{ \frac{G(k^*)}{g'(k^*)} + \frac{G(-\overline{k^*})}{g'(-\overline{k^*})}\right\} \pm 2\pi i \sum_{n=1}^\infty \frac{G(\pm i \beta_n)}{g'(\pm i \beta_n)} \right], \notag \\
&= F^2 i \left[ \pm 2i \Im \left\{ \frac{G(k^*)}{g'(k^*)} \right\} \pm \sum_{n=0}^\infty \frac{G(\pm i \beta_n)}{g'(\pm i \beta_n)}\right], \label{respart}
\end{align}

%Then for solutions in Region I and for $\phi > 0$, we have
%\begin{multline*}
%\theta = \frac{F^2}{2\pi} \left[ -2 \pi i
%\text{Res}\left\{ \frac{G(k)}{g(k)}, k = k_0 \right\} -2 \pi i
%\text{Res}\left\{ \frac{G(k)}{g(k)}, k = -k_0 \right\} \right. \\
%\left. - 2\pi i \sum_{n=1}^\infty \text{Res} \left\{ \frac{G(k)}{g(k)}, k =
%-i\beta_n
%\right\} \right].
%\end{multline*}
%
%\noindent The poles are simple roots of $g(k)$ so using Properties
%(\ref{eq:gclin_gprop1})-(\ref{eq:gclin_gprop2}), we can write
%
%\begin{align} 
%\theta &= -F^2i \left[ \frac{G(k_0)}{g'(k_0)} +
%\frac{G(-k_0)}{g'(k_0)}\right] - F^2i
%\sum_{n=1}^\infty \frac{G(-i\beta_n)}{g'(-i\beta_n)} \notag \\
%&= -F^2i \left[ \frac{k_0 e^{-ik_0\phi}}{g'(k_0)\cosh (\pi k_0)}
%- \frac{k_0 e^{ik_0\phi}}{g'(k_0)\cosh (\pi k_0)} \right] - F^2i
%\sum_{n=1}^\infty \frac{(-i\beta_n)e^{-\beta_n\phi}}{\cos(\pi \beta_n) g'(i\beta_n)} \notag \\
%&= F^2 \left[ \frac{-2k_0}{g'(k_0) \cosh (\pi k_0)}\right] \sin k_0 \phi
%-F^2 \sum_{n=1}^\infty \left[ \frac{\beta_n}{\cos(\pi \beta_n)
%g'(i\beta_n)}\right]
%e^{-\beta_n \phi}. \label{eq:gclin_thetafin_down}
%\end{align}
%
%\noindent Region I solutions with $\phi < 0$ are also handled similarly,
%giving
%\begin{equation} \label{eq:gclin_thetafin_up}
%\theta
%= F^2\left[ \frac{2k_1}{g'(k_1) \cosh (\pi k_1)}\right] \sin k_1 \phi
%- F^2 \sum_{n=1}^\infty \left[ \frac{\beta_n}{\cos(\pi \beta_n)
%g'(i\beta_n)}\right]
%e^{\beta_n \phi}.
%\end{equation}

\noindent where we have used the properties of $g$ and $G$  to conclude that $\frac{G(-\overline{k})}{g(-\overline{k})} = -\overline{\frac{G(k)}{g(k)}}$. From \eqref{respart},  the solution can then be expressed as
\begin{multline} \label{eq:gclin_thetafin_complex}
 \theta_\pm
= F^2\Biggl[ \frac{\pm 2 |k^*| e^{\Im(k^*)\phi}}{|g'(k^*) \cosh (\pi
k^*)|}\Biggr] \sin \Biggl( \Re(k^*) \phi - \text{Arg}(k^*) + \text{Arg}[g'(k^*) \cosh (\pi k^*)]
\Biggr) \\
- F^2 \sum_{n=1}^\infty \left[ \frac{\beta_n}{\cos(\pi \beta_n)
g'(i\beta_n)}\right]
e^{-\beta_n |\phi|},
\end{multline}

\noindent and the $\pm$ signs in (\ref{eq:gclin_thetafin_complex}) are associated with $\phi \lessgtr 0$, respectively. Again, recall that for gravity $(-)$ waves, we set $k^* = k_0$ while for capillary $(+)$ waves, we set $k^* = k_1$. Note that from expression \eqref{eq:gclin_thetafin_complex} for
$\theta$, the (re-scaled) free-surface can then be computed from
\[
 y \sim \delta \int^x \theta(s) \ ds,
\]

\noindent and moreover, in the limit that $\phi \to \pm \infty$, the contributions from $k = \mp i\beta_n$ of (\ref{eq:gclin_thetafin_complex}) tend to zero, leaving only the the contributions from the residues at  $k = \pm k_0, \pm k_1$. 

\subsection{Taking the Low-Froude and low-Bond limits} \label{sec:gclin_lowFB}

\noindent The key result of the previous section, eqn (\ref{eq:gclin_thetafin_complex}), provides the form of the
gravity-capillary waves over a small step. However, we are more interested in the
nature of this approximation as the Froude and Bond numbers tend to zero, so
we let $F = \beta \epsilon$ and $B = \beta \tau \epsilon^2$, where
$\epsilon \ll 1$. This particular scaling for $F$ and $B$ is chosen so that a
power of $\epsilon$ is associated with each derivative of the dependent
variables of (\ref{eq:gclin_dyn_a})---thus $\epsilon$ for the gradient $\nabla \phi$,
and $\epsilon^2$ for the curvature $\kappa$. The reason for this choice will
become clear later when we perform the exponential asymptotic analysis. 
 
With this choice for the Froude and Bond numbers, a balance of terms in eqn
(\ref{eq:gclin_gk}) tells us that $k = \mathcal{O}(1/\epsilon)$. Since $\tanh
(\pi k) \sim 1$ for $\Re(k) > 0$, we can write the dispersion relation \eqref{eq:gclin_gk} as
\[
 \beta \tau (\epsilon k)^2 - \beta(\epsilon k) + 1 \sim 0,
\]

\noindent where we only consider the roots with $\Re(k) > 0$. Then we
have the fact that
\begin{equation} \label{eq:gclin_kep}
k_\pm \sim \frac{D_\pm}{\epsilon},
\end{equation}

\noindent where $D_\pm$ is defined by
\begin{equation} \label{eq:gclin_D}
 D_\pm \equiv \frac{1 \pm \sqrt{\Delta}}{2 \tau},
\end{equation}

\noindent with $\Delta$ defined by
\begin{equation} \label{eq:gclin_Delta}
 \Delta \equiv 1 - \frac{4\tau}{\beta}.
\end{equation}

\noindent The negative sign of eqn (\ref{eq:gclin_kep}) corresponds to gravity waves ($k_{-} = k_0$) and the positive sign to capillary waves ($k_+ = k_1$), and the condition that both wavenumbers are real is  equivalent to $\Delta > 0$ or $\tau < \beta/4$. In fact, this also indicates that the critical dispersion curve $F = F_m$ in Figure \ref{fig:gclin_gctotal} and Figure \ref{fig:gclin_disproots} tends to the curve $B = F^2/4$ as $F \to 0$. Now, using $k_\pm$ (\ref{eq:gclin_kep}) and $D_\pm$ (\ref{eq:gclin_D}) in $g'(k)$ \eqref{eq:gclin_gk}, we can show that
\begin{equation}
 g'(k_\pm) \sim \mp \beta \epsilon \sqrt{\Delta},
\end{equation}

\noindent and also that
\begin{equation} \label{eq:gclin_cosh}
 \cosh (\pi k_\pm) \sim \cosh \left( -\frac{\pi}{\epsilon} D_\pm \right) =
\frac{1}{2}
\left(e^{-\pi D_\pm/\epsilon} + e^{\pi D_\pm/\epsilon}\right) \sim
\frac{e^{\pi D_\pm/\epsilon}}{2}.
\end{equation}

\noindent Returning to the form of the linear solution (\ref{eq:gclin_thetafin_complex}), notice that the terms in the infinite sum are exponentially small as $|\phi| \to \infty$. Certainly for the case where $\Im(k^*) = 0$, the contribution from the sinusoidal waves remains in the far field, but it is still unclear at the moment what role the infinite sum plays, particularly when $\phi = \mathcal{O}(1)$. This will be clarified in the exponential asymptotics to come. We now use (\ref{eq:gclin_kep})--(\ref{eq:gclin_cosh}), and isolate the wave component of the solution (\ref{eq:gclin_thetafin_complex}) in the low-Froude, low-Bond limit. As $|\phi| \to \infty$, the amplitude of the waves (in $\theta$) is shown to be 
\begin{equation} \label{eq:gclin_finaltheta_fourier}
\text{Amplitude of $\theta_\text{exp}$} \sim \frac{4\delta}{\epsilon}\left\lvert
\frac{D_\pm}{\sqrt{\Delta}} \right\rvert \exp \Biggl[
-\frac{\pi\Re(D_\pm)}{\epsilon} - \frac{\lvert \Im(D_\pm)\phi \rvert}{\epsilon} \Biggr] 
%\\ \times \sin \Biggl\{ -\frac{\Re(D_\pm)\phi}{\epsilon} -
%\frac{\pi\Im(D_\pm)}{\epsilon} + \text{Arg}(D_\pm) -
%\frac{\text{Arg}(\Delta)}{2} \Biggr\},
\end{equation}

\noindent where we have multiplied the first-order approximation by $\delta$. Notice that in this final result, we have inserted additional absolute-value signs for $\Im(D_\pm) \phi = \pm \sqrt{\Delta} \phi$. This is in order to dispel the ambiguity of the sign of $\sqrt{\Delta}$ when $\Delta < 0$ and in order to enforce the condition that the waves decay as $\phi \to \pm \infty$ [indeed, recall the restrictions of $k^*$ in \eqref{respart} and \eqref{eq:gclin_thetafin_complex}]. 

This completes our determination of the gravity-capillary waves, and we now move on to retrieve a similar result using exponential asymptotics.

\section{Exponential asymptotics} \label{sec:gclin_expasym}

\noindent In the classical linearisation method of the previous section, we
derived the small-step formula (\ref{eq:gclin_thetafin_complex}) for $F,
B = \mathcal{O}(1)$, only to afterwards take the low-Froude, low-Bond
limits. In the method of exponential asymptotics, we will
instead take the $F, B \to 0$ limit \emph{immediately} after the
initial linearisation. 

As explained in our overview of the methodology, we want to complexify the
free boundary variables ($\xi$, $\phi$, $q$, $\theta$), which is
simply equivalent to replacing the variables by their (2D) complex analogue.
Analytically continuing the boundary integral
(\ref{eq:gclin_bdint_re}) into the upper-half $\xi$-plane gives,
\begin{equation} \label{eq:gclin_bdint_com}
 \log{q} - i\theta =   -\frac{1}{\pi}\int_{-\infty}^0
\frac{\theta(\xi')}{\xi'-\zeta} \ d\xi' +  \mathscr{H}\theta(\zeta),
\end{equation} 

\noindent where we have relabeled $\xi \mapsto \zeta$ and $\mathscr{H}$ denotes a Hilbert transform operator on the free-surface, $\xi' \geq 0$:
 \begin{equation}
 \mathscr{H}\theta(\zeta) =  -\frac{1}{\pi}\int_{0}^\infty
\frac{\theta(\xi')}{\xi'-\zeta} \ d\xi'.
 \end{equation}

\noindent Complexifying the dynamic condition (\ref{eq:gclin_dyn_re}) also gives
\begin{equation} \label{eq:gclin_dyn_com}
 \beta \epsilon \left[ q^2 \frac{dq}{dw} \right]
- \beta \tau \epsilon^2 \left[ q^2 \frac{d^2\theta}{dw^2} +
 q\frac{dq}{dw}\frac{d\theta}{dw} \right] =
- \sin\theta,
\end{equation}

\noindent with a relabeling of $\phi \mapsto w$, and where we have used the
scalings for the Froude and Bond numbers given in \S\ref{sec:gclin_lowFB}.
The expressions in (\ref{eq:gclin_bdint_com})--(\ref{eq:gclin_dyn_com}) then
provide an equation for the complexified free-surface, and this was illustrated
using the perpendicular plane in Figure \ref{fig:gclin_method}. At the end of the analysis, we must also remember to add any solutions from analytic continuation into the \emph{lower}-half $\zeta$-plane.

Assuming that the step is small, and letting $b = 1+\overline{\delta}$, where
$\overline{\delta} \ll 1$, then linearising around the uniform stream gives
\begin{equation} \label{eq:gclin_exp_linexp}
 q = 1 + \overline{\delta} q' + \mathcal{O}(\overline{\delta}^2) \text{\qquad
and \qquad}
 \theta = \overline{\delta} \theta' + \mathcal{O}(\overline{\delta}^2).
\end{equation}

\noindent Then (\ref{eq:gclin_bdint_com}) gives to leading order in
$\overline{\delta}$ (dropping primes)
\begin{equation}
q - i\theta = \frac{1}{2(\zeta + 1)} + \mathscr{H}\theta(\zeta).
\label{eq:gclin_exp_bdint}
\end{equation}

\noindent The dynamic condition in (\ref{eq:gclin_dyn_re}) gives to
leading order in $\overline{\delta}$ (also dropping primes)
\[
\beta\epsilon \frac{dq}{dw} - \beta\tau\epsilon^2 \frac{d^2\theta}{dw^2} =
-\theta,
\]

\noindent and using the fact that
\[
 \frac{d}{dw} = -\zeta \frac{d}{d\zeta}, \qquad \frac{d^2}{dw^2} = \zeta
\frac{d}{d\zeta} + \zeta^2 \frac{d^2}{d\zeta^2},
\]

\noindent the linearised dynamic condition becomes 
\begin{equation}
\beta \epsilon \zeta \frac{dq}{d\zeta} + \beta \tau \epsilon^2 \left[
\zeta^2 \frac{d^2\theta}{d\zeta^2} + \zeta \frac{d\theta}{d\zeta}\right] =
\theta. \label{eq:gclin_exp_dyn}
\end{equation}

\subsection{Outer problem} \label{sec:gclin_outer}

\noindent We first write the solutions of eqns (\ref{eq:gclin_exp_bdint})
and (\ref{eq:gclin_exp_dyn}) as a regular asymptotic expansion: 
\begin{equation}
 \theta \sim \sum_{n=0}^\infty \epsilon^n \theta_n 
 \text{\ \quad and \ \quad} 
 q \sim \sum_{n=0}^\infty \epsilon^n q_n, \label{eq:gclin_series} 
\end{equation}

\noindent which gives at leading order,
\begin{eqnarray} 
 \theta_0 &=& 0, \\
 q_0 &=& \frac{1}{2(\zeta + 1)}, \label{eq:gclin_O1}
\end{eqnarray}

\noindent while at $\mathcal{O}(\epsilon^n$) for $n\geq 1$, we have
\begin{gather}
 q_n = i\theta_n + \mathscr{H}[\theta_n(\zeta)] \label{eq:gclin_bdint_Oen} \\
\beta \zeta \frac{dq_{n-1}}{d\zeta} + \beta \tau \left[
\zeta^2 \frac{d^2\theta_{n-2}}{d\zeta^2} + \zeta \frac{d\theta_{n-2}}{d\zeta}
\right] = \theta_n. \label{eq:gclin_bern_Oen}
\end{gather}

\noindent Now the leading-order solution $q_0$ in (\ref{eq:gclin_O1})
has a singularity in the analytic continuation of the free-surface, $\zeta =
-1,$ which corresponds to the corner of the linearised step. But we see by eqns
(\ref{eq:gclin_bdint_Oen})--(\ref{eq:gclin_bern_Oen}) that at each subsequent
order, $q_n$ is determined by the first and second derivatives of $q_{n-1}$ and
$q_{n-2}$, respectively. Thus, each subsequent order must add to the power of
the early singularity, so that in the limit that $n \to \infty$, the effect of this
early singularity dominates the behaviour of the late-order terms; $q_n$, then, 
will diverge like a factorial over power:
\begin{equation} \label{eq:gclin_ansatz}
q_n \sim \frac{Q\Gamma(n+\gamma)}{\chi^{n+\gamma}},
\end{equation}

\noindent with $\gamma$ constant, $Q$ and $\chi$ as functions of $\zeta$,
and with $\chi(-1) = 0$; of course, the same process leads to a
similar expression for $\theta_n$:
\begin{equation} \label{eq:gclin_theta_ansatz}
\theta_n \sim \frac{\Theta\Gamma(n+\gamma)}{\chi^{n+\gamma}}.
\end{equation}

\noindent At this point, we can profit from yet another simplification: as
$n\to\infty$, and when $\zeta$ is taken near the Stokes line (which we will soon
derive), the integral in (\ref{eq:gclin_bdint_Oen}) is found to be exponentially
subdominant to the other terms of the same equation. Essentially, this occurs
because $|\chi|$ is smallest near the singularity, $\zeta = -1$, and grows along
the Stokes line as it tends towards the free-surface, upon which the integral is
evaluated. This simplification was used in \cite{chapman_2002, chapman_2006},
was discussed in detail for the more interesting case of ship waves in
\cite{trinh_1hull}, and was also rigorously justified in the
context of the Saffman-Taylor viscous fingering problem by \cite{xie_2003}; in
any case, the assumption can simply be checked \emph{a posteriori} after
obtaining the correct form of $\chi$. From (\ref{eq:gclin_bdint_Oen}) then, we
have 
\begin{equation} \label{eq:gclin_qntn}
 q_n \sim i\theta_n,
\end{equation}

\noindent valid as $n\to \infty$. Using this simplification, the
$\mathcal{O}(\epsilon^n$) terms are then given by (\ref{eq:gclin_bern_Oen}),
with 
\begin{equation} \label{eq:gclin_Oep_simp}
 \beta \zeta \frac{dq_{n-1}}{d\zeta} - \beta \tau i \biggl[ \zeta^2
\frac{d^2 q_{n-2}}{d\zeta^2} + \zeta \frac{dq_{n-2}}{d\zeta}\biggr] \sim -iq_n.
\end{equation}

%%%%% MISTAKE HERE???? 

\noindent Substitution of the ansatz in (\ref{eq:gclin_ansatz}) into (\ref{eq:gclin_Oep_simp}) gives to leading order as $n \to \infty$:
\begin{equation} \label{eq:gclin_chieqn}
 (\beta \tau i \zeta^2 ) \left(\frac{d\chi}{d\zeta}\right)^2 + (\beta \zeta)
\frac{d\chi}{d\zeta} - i = 0,
\end{equation}

\noindent or
\begin{equation}
\frac{d\chi}{d\zeta} = -i \left[ \frac{\beta\pm \sqrt{\beta^2 -
4\beta\tau}}{2\beta \tau} \right] \left( -\frac{1}{\zeta} \right) \Longrightarrow
\frac{d\chi}{d\zeta} = iD_\pm \left(\frac{1}{\zeta} \right),
\end{equation}

\noindent with $D_\pm$ as in (\ref{eq:gclin_D}). Requiring that
$\chi(-1)
= 0$, we then have
\begin{equation} \label{eq:gclin_exp_chi}
 \chi = iD_\pm \log\left( -\zeta\right).
\end{equation}

\noindent (For convenience, we shall sometimes drop the subscript notation of $\pm$ where it is not relevant). Proceeding to next order in (\ref{eq:gclin_Oep_simp}), we also have
\[
\frac{dQ}{d\zeta}\biggl[ \beta\zeta i - 2\beta \tau \zeta^2 \frac{d\chi}{d\zeta}\biggr]
= Q\biggl[\beta \tau \zeta^2 \frac{d^2\chi}{d\zeta^2} +  \beta \tau \zeta \frac{d\chi}{d\zeta} \biggr]
\]

\noindent and substituting the equation for $\chi$ in (\ref{eq:gclin_exp_chi})
zeros the right-hand side of the above expression; therefore $Q =
\text{constant} = \Lambda$. Finally, to determine the constant $\gamma$, we note
that in order to match the the leading order solution in eqn
(\ref{eq:gclin_O1}) with the ansatz of eqn (\ref{eq:gclin_ansatz}),
we need $\gamma = 1$.

Before we go on to discuss the implications of these late-order terms, let us inspect the singulant, $\chi$. From \cite{dingle_book}, we know that Stokes lines must emerge from the singularity at $\zeta = -1$, and are prescribed by locations where successive terms in the expansions of $\theta$ and $q$ have the same phase, \emph{i.e.} where
\[
 \Im(\chi) = 0 \qquad \text{and} \qquad \Re(\chi) \geq 0.
\]

\noindent From this, we can verify that in the lower-half complex potential plane
(which we identify with points in physical space), the Stokes lines are given by
the curves $w = \mu(s) \in \mathbb{C}$ with
\begin{equation}
\mu(s) = 
 \begin{dcases}
 si & \text{if $\Delta > 0$ and for $s \in \biggl[-\pi, 0\biggr]$} \\ 
 s + i\left[ \frac{\Re(D_\pm)}{\Im(D_\pm)} s - \pi \right] & \text{if $\Delta <
0$ and for $|s| \in \biggl[0, \pi \frac{\Im(D_\pm)}{\Re(D_\pm)}\biggr]$}. \\
 \end{dcases}
\end{equation}

\begin{figure} \centering
% \begin{preview}
% \begin{tikzpicture}
% \node[inner sep=0pt,outer sep=0pt] at (0,0) {\includegraphics[width=1.0\textwidth]
% 	{figpdf/linear_gcsketch1.pdf}};
% \node[scale=0.8] at (-5.6,1.25) {\scshape capillary};
% \node[scale=0.8] at (5.8,1.25) {\scshape gravity};

% \node[scale=0.7] at (-0.70,0.10) {\mycircd{-2pt}{B} $>$ \mycircd{-2pt}{G}};
% \draw[-stealth] (-0.30, 0.30) arc (135:45:0.45);

% \node[scale=0.7] at (0.7,-0.67) {\mycircd{-2pt}{B} $>$ \mycircd{-2pt}{C}};
% \draw[-stealth] (0.28, -0.5) arc (45:135:0.45);
% \end{tikzpicture}
% \end{preview}
\includegraphics[width=1.0\textwidth]{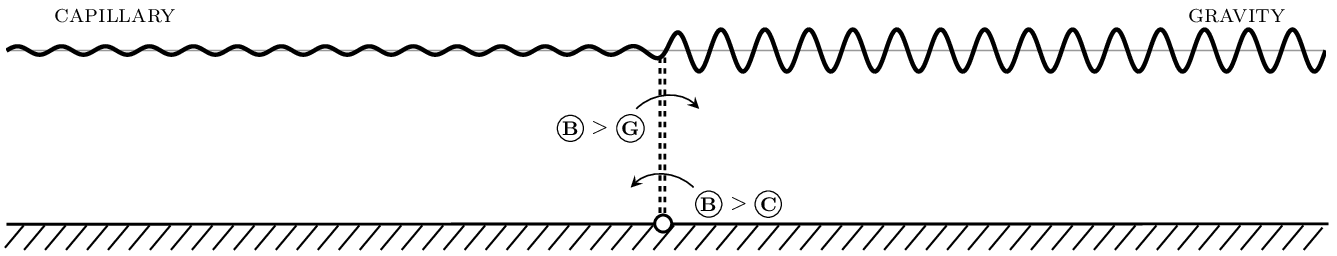}
\caption{%
In Region I with $\Delta > 0$, both capillary and
gravity Stokes lines emerge from $w = -i\pi$ and intersect
the origin. Across this point, we expect a gravity wave to
switch-on downstream and a capillary wave to switch-on
upstream. \label{fig:gclin_stokes1}}
\end{figure}

\noindent Consider the case that $\Delta > 0$. As shown in Figure
\ref{fig:gclin_stokes1}, both the capillary and gravity Stokes lines coalesce
along the imaginary $\psi$ axis, and we would thus expect for waves to switch-on as the free-surface is analytically continued across the origin---gravity waves downstream and capillary waves upstream. This transition is denoted either by \Mycirct{B} $>$ \Mycirct{C} or \Mycirct{B} $>$ \Mycirct{G}, describing the action of the base solution (\ref{eq:gclin_series}) switching on a capillary or gravity wave, respectively. The greater-than sign reminds us that the Stokes line corresponds to the location where the base solution reaches peak exponential dominance over the capillary or gravity wave \citep{dingle_book}. 

\begin{figure}
% \begin{preview}
% \begin{tikzpicture}
% \node[inner sep=0pt,outer sep=0pt] at (0,0) {\includegraphics[width=1.0\linewidth]
% 	{figpdf/linear_gcsketch2.pdf}};
% \node[scale=0.8] at (-5.6,1.25) {\scshape capillary};
% \node[scale=0.8] at (5.8,1.25) {\scshape gravity};

% \node[scale=0.7] at (-2.1,0.15) {\mycircd{-2pt}{B} $>$ \mycircd{-2pt}{C}};
% \draw[stealth-] (-1.95, 0.30) arc (135:50:0.45);

% \node[scale=0.7] at (2.1,0.15) {\mycircd{-2pt}{B} $>$ \mycircd{-2pt}{G}};
% \draw[-stealth] (1.45, 0.45) arc (110:40:0.45);
% \end{tikzpicture}
% \end{preview}
% \begin{center}
 \includegraphics[width=1.0\textwidth]{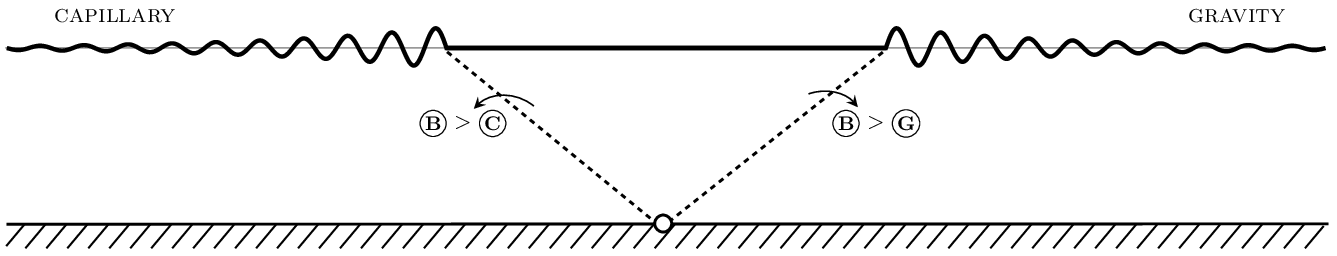}
% \end{center}
\caption{%
In Region II with $\Delta < 0$, the capillary Stokes line moves
upstream while the gravity Stokes line moves downstream. We 
again expect the switching-on of gravity and capillary
waves, but because of the non-zero imaginary component of $\chi$, the
waves are now exponentially decaying as they tend to infinity. 
\label{fig:gclin_stokes2}}
\end{figure}

However, in inferring the behaviour of the Stokes contributions as $\Delta$ is
decreased from $\Delta > 0$ to $\Delta < 0$, we must take great care
in selecting the correct branch of the various square roots if we wish to
continue to identify the same negative sign in $D_\pm$ to gravity waves and the
same positive sign to capillary waves. A much easier mechanism is to see \emph{a
posteriori} which side the exponentials are decaying or growing and to assign
the appropriate wave to the decaying region; we comment on this shortly
in \S\ref{sec:gclin_inner} once we have derived the form of the
exponentials. For the moment, however, it is sufficient to claim that as
$\Delta$ decreases from positive to negative values, the previously coalescent Stokes lines split, with the capillary-associated line moving upstream and the
gravity-associated line moving downstream; this is shown in Figure
\ref{fig:gclin_stokes2}. As the free-surface is analytically continued across
these critical lines, we would again expect small waves to switch on, only now the
waves decay in the far field.

In the next section, we will explicitly show how the Stokes Phenomenon
actually occurs and moreover, we will establish the connection between the resultant waves and the late-order terms of eqns (\ref{eq:gclin_ansatz}) and
(\ref{eq:gclin_theta_ansatz}).

\subsection{Stokes smoothing} \label{sec:gclin_smooth}

\noindent When an asymptotic expansion is truncated haphazardly, the error is
generally algebraically small in $\epsilon$; in order to observe the smooth
switching-on of the waves as the Stokes line is crossed, we must instead
truncate the expansions \emph{optimally}, for which the error is exponentially
small. In fact, for linear differential equations, seeking this remainder term is equivalent to performing a WKB analysis

To begin, we will truncate the asymptotic expansion in $q$,
\begin{equation} \label{eq:gclin_qtrun}
 q = \sum_{n=0}^{\mathcal{N}-1} \epsilon^n q_n + R_\mathcal{N}.
\end{equation}

\noindent Note that the expansions and corresponding expressions for $\theta$
are also similarly done, but the fact that $q_n \sim i\theta_n$ from (\ref{eq:gclin_qntn}) allows us to work with only a single variable. Once the
truncated sum (\ref{eq:gclin_qtrun}) is substituted into eqns
(\ref{eq:gclin_exp_bdint}) and (\ref{eq:gclin_exp_dyn}), we get
\begin{multline} \label{eq:gclin_smooth_firsteq}
 \mathcal{L}(R_\mathcal{N}; \epsilon) + \epsilon^{\mathcal{N}} \Biggl[ \beta \tau i \biggl( \zeta^2
\frac{d^2 q_{\mathcal{N}-2}}{d\zeta^2} + \zeta \frac{dq_{\mathcal{N}-2}}{d\zeta} \biggr) - \beta \zeta \frac{dq_{\mathcal{N}-1}}{d\zeta} \Biggr] \\ +
\epsilon^{\mathcal{N}+1} \beta \tau i \biggl[ \zeta^2 \frac{d^2 q_{\mathcal{N}-1}}{d\zeta^2} + \zeta \frac{q_{\mathcal{N}-1}}{d\zeta}\biggr] = 0,
\end{multline}

\noindent where the linear operator $\mathcal{L}$ is defined by
\begin{equation} \label{eq:gclin_Lop1}
 \mathcal{L}(R_\mathcal{N}; \epsilon)  \equiv \epsilon^2 \beta \tau i \biggl[ \zeta^2 \frac{d^2 R_\mathcal{N}}{d\zeta^2} +
\zeta \frac{dR_\mathcal{N}}{d\zeta} \biggr] - \epsilon \beta \zeta \frac{dR_\mathcal{N}}{d\zeta} -i R_\mathcal{N}.
\end{equation}

\noindent In (\ref{eq:gclin_smooth_firsteq}), we can replace the
$\mathcal{O}(\epsilon^\mathcal{N})$ bracketed quantities with the right-hand side of
eqn (\ref{eq:gclin_Oep_simp}), giving
\begin{equation} \label{eq:gclin_smooth_inhom}
\mathcal{L}(R_\mathcal{N}; \epsilon)  \sim -\epsilon^\mathcal{N} \biggl[ iq_\mathcal{N} + \epsilon\beta\tau
i \Bigl(\zeta^2 \frac{d^2 q_{\mathcal{N}-1}}{d\zeta^2} + \zeta \frac{dq_{\mathcal{N}-1}}{d\zeta} \Bigr)\biggr].
\end{equation}

\noindent It is easily verified that the solution of the homogeneous
equation $\mathcal{L} = 0$ is $R_\mathcal{N} = Qe^{-\chi/\epsilon}$ and so to solve the
inhomogeneous equation, we let
\begin{eqnarray*}
 R_\mathcal{N} &=& \mathcal{S} \left[Q e^{-\chi/\epsilon}\right], \\
 \frac{dR_\mathcal{N}}{d\zeta} &=& \mathcal{S} \frac{d}{d\zeta} \left[ Q e^{-\chi/\epsilon}\right] + \frac{d\mathcal{S}}{d\zeta} \left[
Q e^{-\chi/\epsilon}\right], \\
\frac{d^2 R_\mathcal{N}}{d\zeta^2} &=& \mathcal{S} \frac{d^2}{d\zeta^2} \left[ Q e^{-\chi/\epsilon}\right] + 2\frac{d\mathcal{S}}{d\zeta}
\left[-\frac{\tfrac{d\chi}{d\zeta}}{\epsilon}  Qe^{-\chi/\epsilon} + \frac{dQ}{d\zeta}e^{-\chi/\epsilon}\right]
+ \frac{d^2 \mathcal{S}}{d\zeta^2} Q e^{-\chi/\epsilon},
\end{eqnarray*}

\noindent where now $\mathcal{S}$ describes the \emph{Stokes smoothing}
parameter, which we expect to smoothly vary from zero to a constant across the
Stokes line. When this ansatz for the remainder is substituted into
the left hand-side (LHS) of (\ref{eq:gclin_smooth_inhom}), we are only left with terms
involving derivatives in $\mathcal{S}$ at leading order:
\[
\text{LHS of \eqref{eq:gclin_smooth_inhom}} \sim -\epsilon  \ Qe^{-\chi/\epsilon} \biggl[ \beta
\zeta + 2\beta\tau i \zeta^2 \tfrac{d\chi}{d\zeta} \biggr] \frac{d\mathcal{S}}{d\zeta},
\]

\noindent and switching to differentiation in $\chi$ and using eqn
(\ref{eq:gclin_chieqn}) to change the bracketed quantities, we then have
\begin{equation} \label{eq:gclin_Lop2}
\text{LHS of \eqref{eq:gclin_smooth_inhom}} \sim -\epsilon  \ Qe^{-\chi/\epsilon} \biggl[ i
+ \beta\tau i \zeta^2 \bigl(\tfrac{d\chi}{d\zeta}\bigr)^2 \biggr] \frac{d\mathcal{S}}{d\chi}.
\end{equation}

\noindent Since we are interested in the limit that $\epsilon$ tends to zero, and so the optimal truncation point, $\mathcal{N}$, tends to infinity, we can substitute the late-orders ansatz of (\ref{eq:gclin_ansatz}) into the right hand-side (RHS) of (\ref{eq:gclin_smooth_inhom}), giving
\begin{equation} \label{eq:gclin_Lop3}
\text{RHS of \eqref{eq:gclin_smooth_inhom}}
\sim -\frac{\epsilon^{\mathcal{N}} Q \Gamma(\mathcal{N}+\gamma)}{\chi^{\mathcal{N}+\gamma}} 
\left[ i + \beta\tau i \zeta^2 \bigl(\tfrac{d\chi}{d\zeta}\bigr)^2 \left\{\frac{\epsilon(\mathcal{N}+\gamma+1)}{\chi}\right\}
\right].
\end{equation}

\noindent The optimal truncation point is at $\mathcal{N} \sim \lceil \lvert\chi\rvert/\epsilon \rceil$ (where adjacent terms in the expansion are roughly equal), so we
will let $\chi = re^{i\vartheta}$ and thus set $\mathcal{N} = r/\epsilon + \rho$
where $\rho \in [0, 1)$. Stirling's formula then gives
\[
\Gamma(\mathcal{N}+\gamma) = \Gamma(r/\epsilon + \rho +\gamma) \sim
\sqrt{2\pi}e^{-r/\epsilon} \left(\frac{r^{\mathcal{N}+\gamma}}{\epsilon^\mathcal{N}}\right) 
\frac{r^{-1/2}}{\epsilon^{\gamma-1/2}}. 
\]

\noindent Changing the derivatives to
\[
 \frac{d}{d\chi} = -\frac{ie^{-i\vartheta}}{r} \frac{d}{d\vartheta}
\]

\noindent gives from (\ref{eq:gclin_Lop2}) and (\ref{eq:gclin_Lop3}),
\[
\frac{d\mathcal{S}}{d\vartheta} \biggl[ i + \beta\tau i \zeta^2 \bigl(\tfrac{d\chi}{d\zeta}\bigr)^2
\biggr]
\sim 
\frac{\sqrt{2\pi r}}{\epsilon^{\gamma+1/2}}
\left[ i + \beta\tau i \zeta^2 \bigl(\tfrac{d\chi}{d\zeta}\bigr)^2 \frac{1}{e^{i\vartheta}} \right]
i \left(e^{-i\vartheta}\right)^{\frac{r}{\epsilon}+\rho+\gamma}
e^{\frac{r}{\epsilon}e^{i\vartheta}}e^{-\frac{r}{\epsilon}} e^{i\vartheta}.
\]

\noindent The sum of the exponential factors on the right is exponentially
small, except near the Stokes line $\vartheta = 0$. The critical scaling occurs
with $\vartheta = \sqrt{\epsilon}\overline{\vartheta}$ and here, 
\begin{multline*}
 \exp \left\{
 -\frac{r}{\epsilon} \biggl(1 - \cos\vartheta\biggr)
 + i\left( \frac{r\sin\vartheta}{\epsilon} - \vartheta\left(\frac{r}{\epsilon} +
\rho + \gamma - 1\right) + \frac{\pi}{2}\right) \right\} 
\\ \sim
\exp \left\{ -\frac{r\overline{\vartheta}^2}{2} + \frac{\pi i}{2} +
\mathcal{O}(\sqrt{\epsilon}) \right\}. 
\end{multline*}

\noindent In total then, we have the local change in the Stokes switching as
\begin{equation} \label{eq:gclinear_erf}
\frac{d\mathcal{S}}{d\overline{\vartheta}} \sim \frac{\sqrt{2\pi r}
i}{\epsilon^\gamma}
\exp \left( -\frac{r\overline{\vartheta}^2}{2}
\right).
\end{equation}

\noindent Integrating this expression across the Stokes line from $\overline{\vartheta} = \infty$ (upstream) to $\overline{\vartheta} = -\infty$ (downstream), we find that the jump in the remainder is
\begin{equation} \label{eq:gclinear_Sjump}
 \biggl[ R_\mathcal{N} \biggr]_{\text{upstream}}^{\text{downstream}} \sim -\frac{2\pi i Q}{\epsilon^\gamma} e^{-\chi/\epsilon}
\end{equation}

\noindent and thus across the Stokes line, in a direction from upstream to downstream, the terms
\begin{eqnarray} 
q^{\pm}_\text{exp} &\sim& \frac{4\pi}{\epsilon^\gamma}\Im\left[Q
e^{-\chi_\pm/\epsilon}\right], \label{eq:gclin_qexp} \\
 \theta^\pm_\text{exp} &\sim& \frac{4\pi}{\epsilon^\gamma}\Im\left[\Theta
e^{-\chi_\pm/\epsilon}\right], \label{eq:gclin_thetaexp}
\end{eqnarray}

\noindent are switched on and where, to derive the above expressions, we have
added the contribution from analytic continuation into the lower-half plane
(or simply, the complex conjugate of the contribution from the upper-half
plane). Formulae (\ref{eq:gclin_qexp}) and (\ref{eq:gclin_thetaexp}) thus
provide us with the important connection between the exponentially small waves
and the late-order terms (\ref{eq:gclin_ansatz}). The apparent jump in the remainder, as given by (\ref{eq:gclinear_Sjump}), is computed using an error function (\ref{eq:gclinear_erf}); like the concepts of factorial over power divergence and optimal truncation, this smoothed remainder is another `universal' property of singular problems (see \citealt{berry_1989b, berry_1989}) 

In terms of $\theta$, we will denote the Stokes switching procedure as
\begin{align*}
\text{\small Base series switches-on capillary wave:} && \sum_{n} \epsilon^n \theta_n &\Swtch{}{B}{C}
\sum_{n} \epsilon^n \theta_n + \theta^{+}_\text{exp} \\
\text{\small Base series switches-on gravity wave:} && \sum_{n} \epsilon^n \theta_n &\Swtch{}{B}{G} 
\sum_{n} \epsilon^n \theta_n - \theta^{-}_\text{exp}, 
\end{align*}

\noindent either indicating that the base series has switched on a capillary wave, \Mycirct{B} $>$ \Mycirct{C}, or that the base series has switched on a gravity wave, \Mycirct{B} $>$ \Mycirct{G}. Note that because the Stokes smoothing proceeded from $\overline{\vartheta} = -\infty$ to $\overline{\vartheta} = \infty$, across the Stokes line in the direction of downstream-to-upstream, we then add the negation of the jump corresponding to the gravity wave.

We can finally return to our assertion in \S\ref{sec:gclin_outer} about
the selection of the branches of the various square roots and their association
to the gravity or capillary waves. We first note from the expression for $\chi$ in \eqref{eq:gclin_exp_chi} that when $\zeta \in \mathbb{R}^+$, 
\begin{align} 
 \chi_\pm &= \Bigl[ \pi \Re(D_\pm) - \Im(D_\pm)
\log \zeta \Bigr] + i \Bigl[ \pi \Im(D_\pm) + \Re(D_\pm) \log \zeta \Bigr] \notag \\
&= \Bigl[ \pi \Re(D_\pm) + \Im(D_\pm) \phi \Bigr] + i \Bigl[ \pi \Im(D_\pm) - \Re(D_\pm) \phi \Bigr],\label{chiagain}
\end{align}

\noindent using $\log \zeta = -\phi$. Since the waves are given by (\ref{eq:gclin_thetaexp}), and in particular, their amplitudes are scaled by $e^{-\Re(\chi_\pm)/\epsilon}$, then when $\Delta < 0$, in order for the waves to decay as $\phi \to \infty$ we need $\Im(D_{-}) = -\sqrt{\Delta}/(2\tau)$ to be positive. Thus for $\Delta < 0$, we want $\sqrt{\Delta} = -i\sqrt{|\Delta|}$ for gravity waves. A similar argument holds for assigning $\sqrt{\Delta} = i\sqrt{|\Delta|}$ for the case of capillary waves. With these subtle details in mind, we see that the Stokes
lines in Figure \ref{fig:gclin_stokes2} with $\chi_+$ and $\chi_-$ are indeed
associated with capillary and gravity waves, respectively.

%Examining eqn (\ref{eq:gclin_thetaexp}), we
%see that since we need gravity waves to decay downstream, then for $\Delta < 0$,
%in order to continue identifying the negative sign of $D_\pm$ to gravity waves,
%we need
%\[
%\Re(\chi_{-}) = -\Im(D_{-}) \log\left(\zeta\right) + \pi \Re(D_{-})
%\]
%
%\noindent to be \emph{growing} as $\zeta \to 0^+$. This implies
%that we need $\Im(D_{-})$ to be positive. Thus for $\Delta < 0$, we want the
%secondary branch of the square root $\sqrt{\Delta}$, in order to continue 
%identifying $\chi_{-}$ with gravity waves and $\chi_{+}$ with capillary waves.
%With these subtle details in mind, we see that the Stokes
%lines in Figure \ref{fig:gclin_stokes2} with $\chi_+$ and $\chi_-$ are indeed
%associated with capillary and gravity waves, respectively.

\subsection{Determining $\Lambda$ and the waves} \label{sec:gclin_inner}

\noindent The final step is to determine the unknown pre-factor $Q =
\Lambda$  of eqn (\ref{eq:gclin_qexp}); to do this, we solve for the
leading-order solution near the singularity $\zeta = -a$ and match with the
late-orders expression of eqn (\ref{eq:gclin_ansatz}), which is only valid
far away from the singularity. 

Near the singularity, a similar argument to the one used in deriving (\ref{eq:gclin_qntn}) can be applied to simplify the problem. From (\ref{eq:gclin_exp_bdint}), we would like to evaluate
\begin{equation} \label{eq:gclin_innerbd}
 q_{\textrm{inner}}(w) - i\theta_{\textrm{inner}}(w) =
\frac{1}{2(\zeta+1)} + \mathscr{H}[\theta_\textrm{outer}(\zeta)]
\end{equation}

\noindent as $\zeta \to -1$, where the indices help to remind us where the
functions $q$ and $\theta$ are being evaluated. The left-hand side is evaluated
\emph{near} the singularity and thus involves the exact expressions for $q(w)$
and $\theta(w)$ in the inner limit. However, the integrand on the right hand
side is integrated over the free-surface, far away from the singularity, thus it
involves only the \emph{outer} expansion of $\theta(w)$. But we know that
substituting the outer expansion \eqref{eq:gclin_series} into the right-hand side of (\ref{eq:gclin_innerbd}) leads to
\[ 
 q_{\textrm{inner}}(w) - i\theta_{\textrm{inner}}(w) = q_0 +
\epsilon \mathscr{H}[\theta_1(\zeta)] + \epsilon^2 \mathscr{H}[\theta_2(\zeta)] + \mathcal{O}(\epsilon^3).
\] 
 
\noindent Since the integrals are all finite for $\zeta$ off the axis, the expansion on the right continues to be well-ordered near the singularity. Thus the leading-order inner solution satisfies
\[
 q - i\theta \sim \frac{1}{2(\zeta + 1)},
 \]

\noindent so that
\begin{align*}
 \theta &\sim \frac{i}{2(\zeta + 1)^{\phantom{2}}} - iq, \\
 \frac{d\theta}{d\zeta} &\sim \frac{-i}{2(\zeta +1)^2} -
 i\frac{dq}{d\zeta}, \\
 \frac{d^2\theta}{d\zeta^2} &\sim \frac{i}{\phantom{2}(\zeta +1)^3} -
 i\frac{d^2q}{d\zeta^2}.
\end{align*}

\noindent Substituting these expressions for $\theta$ into (\ref{eq:gclin_exp_dyn}) and simplifying gives
\begin{equation} \label{eq:gclin_inner_q}
\beta \epsilon \zeta \frac{dq}{d\zeta} + \beta \tau \epsilon^2 \Biggl[
-i\zeta^2 \frac{d^2q}{d\zeta^2} - i\zeta \frac{dq}{d\zeta} -
\frac{i\zeta}{2(\zeta+1)^2} + \frac{i\zeta^2}{(\zeta+1)^3}\Biggr] = 
\frac{i}{2(\zeta + 1)} - iq.
\end{equation}

\noindent The scalings for the inner region are given by setting $\epsilon^n q_n \sim q_0$ with \eqref{eq:gclin_series} and \eqref{eq:gclin_ansatz}. This gives $\zeta + 1 = i\epsilon \eta$ and $q = -i\overline{q}/(2\epsilon)$ for inner variables $\eta$ and $\overline{q}$. Using
\[
 \frac{d}{d\zeta} = \frac{1}{i\epsilon}\frac{d}{d\eta},
\]

\noindent and primes to denote differentiation in $\eta$, we then have from eqn (\ref{eq:gclin_inner_q}):
\begin{multline*} 
 \beta \epsilon (-1) \left(\frac{1}{i\epsilon}\right)
   \left(\frac{-i}{2\epsilon}\right) \overline{q}' 
+ \beta \tau \epsilon^2 \left[ -i(-1)^2 \left( \frac{1}{i\epsilon} \right)^2
    \left(\frac{-i}{2\epsilon}\right) \overline{q }'' 
- i(-1)\left(\frac{1}{i\epsilon}\right)
    \left(\frac{-i}{2\epsilon}\right)\overline{q}' \right. \\
\left. - \frac{i(-1)}{2(i\epsilon \eta)^2} + \frac{i(-1)^2}{(i\epsilon
    \eta)^3}\right] = \frac{i}{2}\frac{1}{(i\epsilon \eta)} - i \left(\frac{-i}{2\epsilon}\right)
    \overline{q},
\end{multline*}

\noindent and after simplifying, this gives
\[
\frac{\beta}{2\epsilon}\overline{q}' + \frac{\beta\tau}{\epsilon} \left[
\frac{1}{2} \overline{q}'' - \frac{1}{\eta^3} + \mathcal{O}(\epsilon) \right] =
\frac{1}{2\epsilon \eta} - \frac{1}{2\epsilon} \overline{q},
\]

\noindent so that the equation for the leading-order inner solution is 
\begin{equation} \label{eq:gclin_innereq}
 \beta \overline{q}' + \beta \tau \left[ \overline{q}'' -
\frac{2}{\eta^3} \right] = \frac{1}{\eta} - \overline{q}.
\end{equation}

\noindent By examining the form of the outer ansatz (\ref{eq:gclin_ansatz}), we expect that in the limit that the inner solution tends to the outer region, it must also diverge like a factorial over power. If we write
\[
 \overline{q} = \sum_{n=0}^\infty \frac{A_n\Gamma(n+1)}{\eta^{n+1}},
\]

\noindent as $\eta \to \infty$, this gives from (\ref{eq:gclin_innereq}),
\begin{multline*}
 -\sum_{n=1}^\infty \frac{A_{n-1}(n)\Gamma(n)}{\eta^{n+1}} + \beta\tau \left[
 \sum_{n=2}^\infty \frac{A_{n-2}(n+1)n\Gamma(n+1)}{\eta^{n+1}} -
 \frac{2}{\eta^3} \right] = 
 \frac{1}{\eta} - \sum_{n=0}^\infty \frac{A_n \Gamma(n+1)}{\eta^{n+1}}.
\end{multline*}

\noindent In the limit that $\eta \to \infty$, the first three orders give
\begin{subequations}
\begin{align}
 1 - A_0 &= 0 &\Rightarrow&& A_0 &= 1, \label{eq:gclin_A0} \\
 -\beta A_0 \Gamma(1) &= -A_1 \Gamma(2) &\Rightarrow&& A_1 &= \beta,
\label{eq:gclin_A1} \\
 -2\beta A_1 + \beta \tau \left[ 2A_0 - 2 \right] &= -2 A_2 &\Rightarrow&&
A_2 &= \beta^2, \label{eq:gclin_A2}
\end{align}
\end{subequations}

\noindent while for $n \geq 3$, 
\begin{equation} \label{eq:gclin_An}
 A_n = \beta \left( A_{n-1} - \tau A_{n-2} \right).
\end{equation}

\noindent Using $D_\pm$ from (\ref{eq:gclin_D}), we can show that
the solution of the recurrence relation in (\ref{eq:gclin_An}) is 
\begin{equation} \label{eq:gclin_Ansol}
 A_n = \frac{1}{\sqrt{\Delta}}\left\{ 
 \left(\frac{1}{D_-}\right)^n - \left(\frac{1}{D_+}\right)^n
 \right\}
\end{equation}

\noindent with $\Delta$ given by (\ref{eq:gclin_Delta}). In order to
perform the matching between inner and outer expansions, we first note that 
\begin{equation} \label{eq:gclin_chi_inner}
 \chi_\pm = -i D_\pm \log \left( - \zeta \right) \sim iD_\pm
\left(\zeta+1\right) = -D_\pm \epsilon \eta,
\end{equation}

\noindent and then matching is done by applying Van Dyke's \citeyearpar{vandyke_book} rule. The $n^\text{th}$ term of the outer solution (n.t.o),
written in inner variables and re-expanded to one term (1.t.i) is
\begin{equation}
 q \xrightarrow[]{\text{($n$.t.o)}} \epsilon^n q_n \sim \frac{\epsilon^n
\Lambda \Gamma(n+1)}{\chi^{n+1}} \xrightarrow[]{\text{($1$.t.i)}}
\frac{(-1)^{n+1} \Lambda \Gamma(n+1)}{\epsilon D_\pm^{n+1} \eta^{n+1}}
\end{equation}

\noindent where we have used the inner limits of $\chi$ from eqn
(\ref{eq:gclin_chi_inner}). Similarly, the leading-order term of the inner region
(1.t.i),
re-expanded to the $n^\text{th}$ term in the outer limit (n.t.o) has
\begin{equation*}
 q \xrightarrow[]{\text{(1.t.i)}} \frac{-i}{2\epsilon}
\overline{q} \xrightarrow[]{\text{($n$.t.o)}} -\frac{i}{2\epsilon} \frac{A_n
\Gamma(n+1)}{\eta^{n+1}},
\end{equation*}

\noindent and using the exact solution of the recurrence relation in (\ref{eq:gclin_Ansol}) requires
\begin{equation} \label{eq:gclin_lambda}
 \Lambda = \pm \frac{i}{2}\frac{D_\pm}{\sqrt{\Delta}}.
\end{equation}

\noindent This completes our determination of all the components of the high-order terms of (\ref{eq:gclin_ansatz}), as well as the exponentials switched-on across Stokes lines from (\ref{eq:gclin_qexp})--(\ref{eq:gclin_thetaexp}). 

Indeed, from (\ref{eq:gclin_qntn}), we know that $\Theta = -iQ = -i\Lambda$,
and thus from (\ref{eq:gclin_lambda}), we have 
\begin{equation} \label{Theta}
 \Theta = -i\Lambda = \pm \frac{1}{2} \left\lvert \frac{D_{\pm}}{\sqrt{\Delta}}
\right\rvert \exp \left[i\left(\text{Arg}(D_\pm) - \frac{\text{Arg}
(\Delta)}{2}\right)\right].
\end{equation}

\noindent Remembering that we must set $\Im(D_\pm) \phi = -\lvert \Im(D_\pm) \phi\rvert$ in order to ensure the proper branch of $\sqrt{\Delta}$ is chosen, we may now combine the components of $\chi$ \eqref{chiagain} and $\Theta$ \eqref{Theta} into the wave expression for $\theta_\text{exp}$ \eqref{eq:gclin_thetaexp} to find that the amplitude of the waves are
\begin{equation} \label{eq:gclin_finaltheta_exp}
\text{Amplitude of $\theta_\text{exp}$} \sim \frac{2\pi\overline{\delta}}{\epsilon} \left \lvert
\frac{D_\pm}{\sqrt{\Delta}}\right\rvert \exp \Biggl[
-\frac{\pi\Re(D_\pm)}{\epsilon} - \frac{\lvert \Im(D_\pm)\phi \rvert}{\epsilon} \Biggr] 
%\times \sin \Biggl\{ -\frac{\Re(D_\pm)\phi}{\epsilon} -
%\frac{\pi\Im(D_\pm)}{\epsilon} + \text{Arg}(D_\pm) -
%\frac{\text{Arg}(\Delta)}{2} \Biggr\},
\end{equation}

\noindent after having multiplied by $\overline{\delta}$ (since we chose to
drop primes following the expansion of \eqref{eq:gclin_exp_linexp}. A
simple flux argument gives $\pi\overline{\delta} = 2\delta$,
relating the step height in the physical and potential planes, and so we 
see that indeed the waves from the standard Fourier analysis
(\ref{eq:gclin_finaltheta_fourier}) are the same as the waves from the
exponential asymptotics (\ref{eq:gclin_finaltheta_exp}).

\section{Discussion} \label{sec:gclin_discuss}

\noindent \emph{What is the connection between the
treatment of the low-Froude, low-Bond problem using traditional Fourier methods
and our new methods in exponential asymptotics}? 

For both the Fourier method and the exponential asymptotics, the
free-surface waves given in (\ref{eq:gclin_finaltheta_fourier}) and
(\ref{eq:gclin_finaltheta_exp}) are produced by deforming a path of
integration through the relevant singularities---\emph{explicitly} in
Fourier-space for the former, and \emph{implicitly} in the analytically
continued domain for latter. We find, however, that when we use
exponential asymptotics, concepts like the Stokes Phenomenon and optimal
truncation can be clearly interpreted, but if one only sees the problem
through Fourier-tinted glasses, then these equivalent notions remain well
obfuscated.

Consider the case that $\Delta > 0$. The analysis by exponential asymptotics
clearly indicates that the formation of the gravity and capillary waves occurs
upon crossing the point $\phi = 0$ on the free-surface, and this is reflected in
the integration contour in the Fourier plane, which runs in the upper-half
$k$-plane upstream, but switches into the lower-half $k$-plane downstream.
However, for $\Delta < 0$, the exponential asymptotics clearly show that the
capillary and gravity waves only appear on \emph{either side} of the point $\phi
= 0$. In particular, this seems to imply that for a small region near $\phi =
0$, the integration contour within Fourier space does not include the complex
residue contributions of the four poles---\emph{there are no waves after all!}
The connection between the optimal truncation and Stokes smoothing procedure
with the standard linearised Fourier theory as $\epsilon \to 0$ is no longer
clear. 

How do the infinitude of complex poles in the Fourier
plane relate to the optimally truncated asymptotic solution? From
(\ref{eq:gclin_bdint_Oen}), the low-Froude, low-Bond solution has
\begin{equation} \label{eq:gclin_discuss_lowsol}
 \theta = -\beta\epsilon \delta \left[ \frac{\zeta}{(\zeta+1)^2} \right] +
\mathcal{O}(\epsilon^2).
\end{equation}

\noindent This expression must be related to the infinite summation of the residues along the
positive or negative imaginary axis of eqn
(\ref{eq:gclin_thetafin_complex}),
which we write as
\begin{equation} \label{eq:gclin_thetafin2}
 \theta = - \beta \epsilon \delta \sum_{n=0}^\infty \left[
\frac{\beta_n}{\cos(\pi \beta_n) g'(i\beta_n)}\right] e^{-\beta_n \phi} +
\mathcal{O}(e^{-\chi/\epsilon}),
\end{equation}

\noindent having multiplied by the size of the step $\delta$. As $\epsilon
\to 0$, the complex roots $\pm i \beta_n$ can be found by setting $\beta_n = n +
d$ so that from eqn (\ref{eq:gclin_Bn_eqn}) we find
\[
 d + \frac{1}{3}d^3 +
\mathcal{O}(d^5) = \beta^2\epsilon^2 (n +
d)\biggl[1 +
\beta \tau \epsilon^2 (n+d)^2 + \mathcal{O}(\epsilon^4,
d^4)\biggr],
\]

\noindent and $d$ can be expanded as a (regular) perturbation
series in powers of $\epsilon$. Now if we substitute $\beta_n \sim n$ back into
eqn (\ref{eq:gclin_thetafin2}), we find
\[
 \theta = -\beta \epsilon \delta \sum_{n=0}^\infty
(-1)^{n+1} n \zeta^n + \mathcal{O}(\epsilon^2),
\]

\noindent after making the substitution $\zeta = e^{\phi}$. In fact, we may
truncate this sum at $n = \mathcal{N}$ with
\[
\sum_{n=0}^\mathcal{N} (-1)^{n+1} n \zeta^n = \left[ \frac{\zeta}{(\zeta+1)^2}
\right] \biggl\{ 1 - (-\zeta)^\mathcal{N} + \mathcal{N} (-\zeta)^\mathcal{N} + (-1)^{\mathcal{N}}\mathcal{N}
\zeta^{\mathcal{N}+1}\biggr\}.
\]

\noindent Then selecting $\mathcal{N} \sim \log\epsilon/\log \zeta$, we have
\[
\theta = - \beta \epsilon \delta \left[\frac{\zeta}{(\zeta+1)^2} \right] +
\mathcal{O}(\epsilon^2).
\]

\noindent What we have thus shown is that by adding $\mathcal{N}$ of the
residues along the imaginary axis, we can reconstruct the \emph{exact} form of
the leading-order term of the asymptotic approximation
(\ref{eq:gclin_discuss_lowsol}) using only the leading order behaviour of
$\beta_n$. For a general term of the optimally truncated approximation
(\ref{eq:gclin_qtrun}), we would need to also truncate the expansion of each
residue
\[
 \beta_n = \beta_{n}^{(0)} + \epsilon \beta_{n}^{(1)} + \epsilon^2
\beta_{n}^{(2)} + \ldots,
\]

\noindent and then add up a further truncated sum from
eqn (\ref{eq:gclin_thetafin2}) in order to retrieve the precise form. But
because this involves a composite function of many truncated sums, the full
process is likely to be excessively complicated. The exponential asymptotics indicates that the region near the center of the free surface should be wave free, and so we conclude that the infinite sum (\ref{eq:gclin_thetafin2}) must be destructively interfering with the Fourier waves (\ref{eq:gclin_finaltheta_fourier}) when $\phi = \mathcal{O}(1)$.

Ultimately, however, the bridge between Fourier analysis and exponential asymptotics becomes less important as we move on to study the nonlinear problem, where only the latter method is applicable. In the next part \citep{trinh_gcnonlinear}, we will show how exponential asymptotics can be used without first linearising for small obstructions, thus providing an asymptotic approximation valid for $\mathcal{O}(1)$ geometries. There, we shall see that the availability of multiple singularities in the geometry, coupled with the interplay of gravitational and cohesive effects, leads to the discovery of a remarkable set of new gravity-capillary waves.

\appendix
\section{Linear classification of solutions}
\label{sec:gclin_class}

\noindent For the finite-depth flow in a channel of height $\pi$, the dispersion
relation and its derivative can be written as
\begin{eqnarray} 
 g(k) &=& kF^2 - \text{tanh} (\pi k)(Bk^2 + 1), \label{eq:gclin_dispersion} \\
 g'(k) &=& F^2 - \text{sech}^2(\pi k)(Bk^{2}+1) - \tanh(\pi k)(2kB).
\label{eq:gclin_gp}
\end{eqnarray}

\begin{figure}
\begin{minipage}[t]{0.6\textwidth}
\vspace*{0pt}
% \begin{preview}
% \vspace*{0pt}
%  \begin{tikzpicture}  
%    \begin{axis}[xlabel={Froude, $F$}, 
%  		ylabel={Bond, $B$}, 
%  		ylabel style={rotate=90, font=\large},
%  		xlabel style={font=\large},
%  		ymin=0, ymax=0.4,
%  		xmin=0, xmax=1.3,
%  		width=\textwidth, height=8.5cm]
 		
%  	\addplot[smooth, line width=1.5pt] file {plotdata/dispersion.dat};
%  	\addplot[dashed, line width=1.0pt] coordinates {%
%  		(1,0) (1,0.5)};
%  	\addplot[dashed, line width=1.0pt] coordinates {%
%  		(0,1/3) (1,1/3)};
%  	\addplot[mark=*, only marks, %
%  		mark options={%
%  		scale=1.6, fill=gray!15, draw=black, line width=0.8pt}] 
%  		coordinates {(1,0.3333)};
%  	\node[scale=1.2] at (axis cs:0.85,0.05) {I};
%  	\node[scale=1.2] at (axis cs:0.45,0.2) {II};
%  	\node[scale=1.2] at (axis cs:1.15,0.2) {III};
%  	\node[scale=1.0] at (axis cs:0.45,0.35) {$B = 1/3$};
%    \end{axis}
%    \end{tikzpicture}
% \end{preview}
\includegraphics[width=1.0\textwidth]{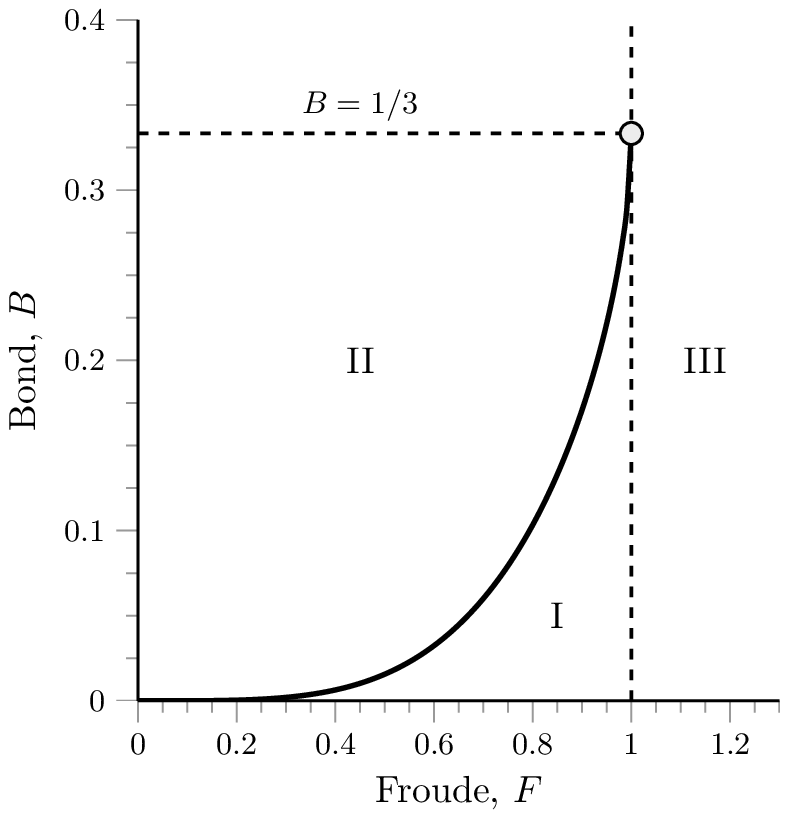}
\end{minipage} \hfill
\begin{minipage}[t]{0.35\textwidth} \centering
\vspace*{12pt}
% \beginpgfgraphicnamed{gclinear_dispersion2}
% \begin{tikzpicture}
% \node at (0,0)
% {\includegraphics[width=\linewidth]{figpdf/dispersionroots.pdf}};
% \node at (0,1.2) {I to III};
% \node at (0,4.2) {I to II};
% \node at (0,-1.6) {III to II};
% \end{tikzpicture}
% \endpgfgraphicnamed
\includegraphics[width=1.0\textwidth]{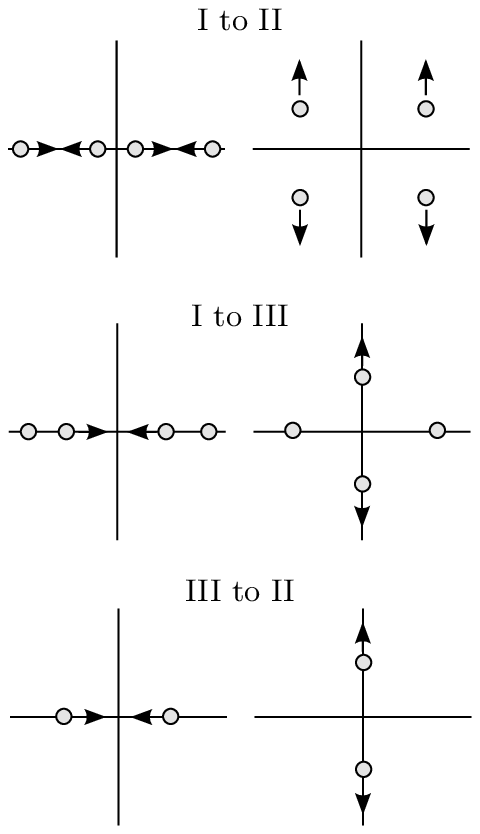}
\end{minipage}
\caption{The three regions in $(F, B)$ space which distinguish the linearised solutions are shown on the left. The transitions between the different regions can be understood by following the movements of the four (possibly repeated) wavenumbers in the complex complex $k$-plane, shown on the right.\label{fig:gclin_disproots}}
\end{figure}

\noindent The nature of the zeros of the dispersion relation provide a
classification of the linearised solutions in the $(F,B)$ plane. For all values
of $F$ and $B$, $k = 0$ is a root of (\ref{eq:gclin_dispersion}), but
it plays no role unless $F = 1$. There are an infinite number of purely imaginary
roots at $k = \pm i\beta_0, \pm i\beta_1, \ldots$. The relevant regions
in the linear classification are shown in Figure \ref{fig:gclin_disproots}.

First, consider Region I of the $(F,B)$ plane. Here, the dispersion relation has four real roots: $k = \pm k_0$ and $k = \pm k_1$ where $0 < k_0 < k_1$; clearly, $k_0$ corresponds to the gravity waves and $k_1$ to the capillary waves. Physically, we would expect capillary waves upstream and gravity waves downstream, and this subcritical region is the one relevant to Rayleigh's \citeyearpar{rayleigh_1883} fishing-pole problem.

As $B \to 0$, $\pm k_1 \to \pm \infty$ and only the gravity waves remain.
Alternatively, if we hold $F$ fixed and increase $B$, then the roots coalesce at
a critical value (corresponding to the dispersion curve $F = F_m$) and for $B$
still larger, they split, forming four complex roots; here, we would expect
localised solitary solutions. Physically, this bifurcation point corresponds to
point where the group velocity of the wave equals the phase velocity, so that
$k_0 = k_1 = k^*$ and critical dispersion curve $F = F_m$ is then defined by
solving $g(k^*) = g'(k^*) = 0$.

Next, suppose we begin again in Region I, hold $B$ fixed, and increase
$F$. Then we find that the two gravity roots coalesce at the origin, and then
split, moving upwards and downwards on the imaginary axis. This bifurcation
describes the more complex phenomena of the exponentially radiating waves of the
fifth-order Korteweg-de Vries eqn.

Finally, if we begin in Region III with $B > 1/3$, and hold $B$ fixed whilst
decreasing $F$, we transition back to Region II. This corresponds to having
the two capillary roots coalescing near the origin, and then splitting onto the
imaginary axis. 

\bibliographystyle{jfm}
%\bibliography{philmaster}
\providecommand{\noopsort}[1]{}

\end{document}